# Negative Effects of Gamification in Education Software: Systematic Mapping and Practitioner Perceptions

Cláuvin Almeida[a], Marcos Kalinowski[a], Anderson Uchôa[b], Bruno Feijó[a]

*[a]Pontifical Catholic University of Rio de Janeiro (PUC-Rio), Rio de Janeiro, Brazil*
*[b]Federal University of Ceará (UFC), Itapajé, Brazil*

## Abstract

**Context:** While most research shows positive effects of gamification, the focus on its adverse effects is considerably smaller and further understanding of these effects is needed. **Objective:** To provide a comprehensive overview on research reporting negative effects of game design elements and to provide insights into the awareness of developers on these effects and into how they could be considered in practice. **Method:** We conducted a systematic mapping study of the negative effects of game design elements on education/learning systems. We also held a focus group discussion with developers of a gamified software, discussing the mapping study results with regard to their awareness and perceptions on the reported negative effects in practice. **Results:** The mapping study revealed 87 papers reporting undesired effects of game design elements. We found that badges, leaderboards, competitions, and points are the game design elements most often reported as causing negative effects. The most cited negative effects were lack of effect, worsened performance, motivational issues, lack of understanding, and irrelevance. The ethical issues of gaming the system and cheating were also often reported. As part of our results, we map the relations between game design elements and the negative effects that they may cause. The focus group revealed that developers were not aware of many of the possible negative effects and that they consider this type of information useful.

*Email addresses:* almeidaclauvin@gmail.com (Cláuvin Almeida),
kalinowski@inf.puc-rio.br (Marcos Kalinowski),
andersonuchoa@ufc.br (AndersonUchôa),
bfeijo@inf.puc-rio.br (Bruno Feijó)

The discussion revealed their agreement on some of those potential negative effects and also some positive counterparts. **Conclusions:** Gamification, when properly applied, can have positive effects on education/learning software. However, gamified software is also prone to generate harmful effects. Revealing and discussing potentially negative effects can help to make more informed decisions considering their trade-off with respect to the expected benefits.

---

## 1. Introduction

There are plenty of digital platforms for education with a massive number of users, as Duolingo, a language teaching service used by 300 million people worldwide. It boasts that it gives opportunities for people to learn new languages, no matter their financial standing [1].

Duolingo and other services worldwide use gamification - applying game-playing elements to non-game contexts [2] that are typically tedious, discouraging, or inefficient - as a strategy to make their objectives more achievable. It's a strategy with a strong presence in education, and other domains [3], representing a market predicted to grow over 30% through 2019-2025, with an expected value of more than 32 billion in 2025 [4]. This context means that a significant demand for gamified software exists, which calls for software engineers to develop them.

Software development is not a trivial task, and it is more complicated in the case of gamified software solutions. These cases require specialized expertise, going beyond what is expected by an average software engineer [5], for instance:

- Effective gamification requires knowledge of human psychology, similar to how serious games require knowledge regarding the subject they deal with. This necessity arises because gamified software aims to change human behavior (see Volkswagen's Fun Theory videos ([6],[7],[8]);

- Software engineers need a good understanding of the game design mechanics used as tools and how they contribute to functional and non-functional requirements;
- Software engineers face the fact that gamified software has a more limited design space and different objectives to focus on than a game [9].

Hence, selecting the correct gamification elements when designing gamified software is strongly related to requirements engineering and can affect the overall project success [10]. Moreover, defects in requirement are the most expensive to fix when found in production [11][12]. Indeed, given that gamification deals with changing human behavior, when gamified software is ill-specified, it may not hit the intended target or even be counterproductive, which can have serious consequences when applied to education.

Education is the main target of the present work for important reasons. For instance, approximately 617 million children and adolescents of primary and lower secondary school age (roughly 55% of the global total) have not reached minimum reading and mathematics proficiency in 2015 [13]. The reasons for this global learning crisis are manifold, such as inequality and poverty, but the poor quality of education is one of the critical causes. In this context, applying gamification to education and learning represents a promising means to allow educators to make learning fun, contextualize learning quickly, speak the language of young people, and directly deal with soft skills, improving education quality. However, gamification is also prone to generate harmful effects, usually unknown to designers and engineers. These unexpected effects happen because current gamification research lacks a critical lens capable of exploring unintended design consequences [3].

In this paper we extend our previous systematic mapping study efforts on adverse, unintended gamification effects in education and learning software (such as modified learning management software, coursewares and digital learning environments) [14]. Our search strategy is driven by broad research questions and uses a hybrid search strategy [15], combining database search with back-ward and forward snowballing. The extension comprises providing more details on the mapping study, complementing the search strategy to gather evidence reported until the end of 2020 (previously 2020 was not completely covered), leading our set of identified studies to increase from 77 to 87. As far as we know, this paper is the most precise, comprehensive, and up-to-date systematic mapping study that organizes evidence regarding the adverse effects of game design elements in gamified education software. Furthermore, we conducted a focus group discussion with developers of a gamified software, discussing the mapping study results with regard to their awareness and perceptions on the reported negative effects.

The mapping study reported several different negative effects, allowing to gather valuable information for software engineers and designers of gamified education/learning software, such as the game design elements that have most often been reported being involved with adverse effects; the most common negative effects on students; the adverse consequences affecting teachers; the relation between game design elements and negative effects; the fields in which research on negative effects has been conducted; and the types of empirical studies conducted to assess the adverse effects.

The study revealed that badges, leaderboards, competitions, and points are the game design elements most often reported as causing negative effects. The most reported negative effects were lack of effect, worsened performance, motivational issues, lack of understanding, and irrelevance. The ethical issues of gaming the system and cheating were also often reported. We also mapped the relations between game design elements and the negative effects that they may cause, so that these effects can be pragmatically considered when designing gamified software.

The focus group revealed that developers were not aware of many of the possible negative effects and that they consider this type of information useful. We report on the discussions held with the practitioners on the use of the identified game design elements and their negative effects. Nevertheless, they recall that this information should be discussed within the context of the gamified software, as in this focus group, to analyze trade-offs between expected positive effects and potential negative ones. Sharing this discussion provides further insights into the usefulness of our mapping study results from the point of view of the practitioners.

The organization of the paper is as follows. Section 2 presents the fundamentals of gamification and gamification effects and related work. We explain our systematic mapping protocol and its execution in Section 3. The results answering the research questions are presented Section 4. In Section 5 we describe the focus group its results. Finally, Sections 6 and 7 discuss the limitations and present the conclusions.

## 2. Background and Related Work

In this section we provide the background on gamification, game design elements and its effects. We also present related work on negative effects, including a comparison to other secondary studies that have been conducted and explaining the differences to ours.

### *2.1. Gamification*

The term "gamification" became known after 2010, but we can find precursors in user interface studies of the 80's [2]. If we use the broader perspective of work's gamification, we find similar concepts even earlier, such as the Soviet Union's experiments to motivate workers of the first half of the 20th century [16]. At the beginning of the '2000s, the American movement of "fun at work" in the academic management literature was also a form of work's gamification [16].

Yu-Kai Chou [17] defines gamification as the craft of deriving fun and engaging elements found typically in games and thoughtfully applying them to real-world or productive activities. He calls this process Human-Focused Design because it concentrates on humans' feelings, motivations, and engagement in the experience.

Unfortunately, the term "gamification" remains inconsistently used, and a general theory of gamification is yet to be developed [18]. Aware of so many different definitions, discrepancies, distinctions, and discretionary delimitations, we use the original concept proposed by Deterding et al. [2]: *Gamification is the use of game design elements in non-game contexts.*

To allow a more precise understanding of this definition, we define game design according to Brathwaite and Schreiber [19], i.e.: "Game design is the process of creating the content and rules of a game. Good game design is the process of creating goals that a player feels motivated to reach and rules that a player must follow as he makes meaningful decisions in pursuit of those goals.".

For game design elements, we use the definition by Deterding et al. [2] "(. . . ) elements that are characteristic to games – (. . . ) that are found in most (but not necessarily all) games, readily associated with games, and found to play a significant role in gameplay." We provide more detail about the game design elements in the next subsection.

### *2.2. Game Design Elements*

*Game Design Elements* (GDEs) represents the basic components of any gamified software [2, 20]. For instance, assigning points to players directly reward user interaction promotes further interactions in the future. Additionally, it might stimulate competitive environment among players, to achieve the highest scores in points. In order to gamify any software, many gamification design frameworks were proposed in the literature [21, 22, 23, 24]. These frameworks tend to present and summarize different types of GDEs. For instance, [21] proposed a framework that relies on six gamification steps, that contains 30 game elements. Conversely, [23] proposed a framework based on seven aspects that may influence gamification, such framework contains 14 GDEs. Despite the difference among the gamification design frameworks, there are a set of GDEs which are commonly described. Table 1 list some of these GDEs.

### *2.3. Gamification Effects*

Some scholars have a negative opinion on gamification. For example, Ian Bogost [30] stated that gamification is primarily an opportunistic marketing strategy. Despite such criticism, research over the years found that gamification does bring benefits when properly used.

Table 1: Description of game design elements

| GDEs | Description |
|---|---|
| Avatar | Visual representation of the player within the game. It can make player feel included and comfortable within the game [2]. |
| Badges/Achievements | Visual representation of achievements the player has earned within the game [21]. Badges can have different functions depending on how they are designed. For instance, they can be used to create a comparison with others or to challenge oneself [25]. |
| Challenges | Elements that represent a set of tasks or actions, to be achieved by the player [2]. The challenges can include time limitation for a certain task (e.g., answering a question in the shortest possible time), participation reinforcement (e.g., complementing the answer of colleagues in a forum discussion), among others [26]. |
| Experience Points (XPs) | Points used as a measure for a player's progression towards their next goal or experience level [26]. |
| Feedback | Consists of alerts about a specific set of player actions, that allows comparison between an actual outcome and a desired outcome [27]. Feedback is intended to provide the users with information about their performance or the status of their actions, which makes it possible to change behavior [28]. |
| Goals | Consist of an objective that the users need to achieve. Frequently, used together with challenges and tasks that involve points [26]. |
| Leaderboards | Consist of a board that ranks players from highest to lowest or lowest to highest based on their scores/achievements/items in a game [29]. It can affect the player's behaviors and outcomes since it increases competition [2, 21]. |
| Points | Unit rewarded to players. They are often linked to levels and can motivate the players by making them perform more actions in order to gain more points [2, 21]. |
| Quizzes | Consist of a set of questions that aim to assess the players' knowledge on a given subject [21, 2, 28]. |
| Rankings | Consist of a player's relative placement in the game [26]. |
| Rewards | Represent a reward the player earns for doing a specific set of activities or after unlocking an achievement [2]. |
| Virtual Currencies | These are in-game items that players can collect and use in a virtual rather than a real way. Players can pay for in-game items or currency or with real money [26]. |

Using an action research design, Putz et al. [31] found that gamification can have a positive effect on students' knowledge retention, independent of age and gender. Positive effects of gamification on enhancing interaction with learning materials and performance on studies were also reported by Klock et al. [32]

Systematic reviews have revealed both positive and negative effects. Zainuddin et al. [33] found evidence that the use of game design elements such as badges, points, trophies, leader boards, avatars, and virtual gifts not only promotes students' extrinsic motivation but also increases their intrinsic value for learning; however, they also reported studies with contradic tory findings. Johnson et al. [34] conducted a systematic literature review of empirical studies on gamification for health and well-being. From the papers identified, the impact of gamified interventions was found to be positive by 59% of the papers reviewed, with effects including empowerment, motivation, health monitoring, and more healthy habits taken. However, 41% - a significant portion of the studies - reported mixed or neutral effects.

Hamari et al. [35] corroborated the point about mixed effects. Most quantitative studies identified in their review concluded positive effects of gamification elements to exist only partially. Also, they observed (through qualitative analysis) that gamification is more manifold than previous studies often assumed. Koivisto and Hamari [36] reached the same conclusion, having found papers pointing to a mixed effect of gamification and a small amount of purely negative results, which they attributed to a possible confirmation bias.

Indeed, gamification manipulates human psychology through game design elements, and it is natural to expect that such manipulation can have adverse effects. One issue occurred at the Disneyland Resort Hotel, California, in 2018. The hotel decided to use leaderboards updated in real-time to stimulate its workers to clean sheets and towels more efficiently. The initiative backfired hard because the competition degenerated the environment's quality, caused extra stress, and increased the number of injuries on the job [37]. Another case happened with Go365, a gamified app, when imposed on public school teachers in West Virginia, forcing them to provide sensitive medical data and have their positions tracked daily. This gamified app's enforcement was the last in a series of events that lead to a wildcat strike in 2018 [37]. Recent research points out that a gamified activity should never be mandatory; otherwise, it loses its fun value and leads to the collapse of play and work value [38].

### *2.4. Related Work on Gamification Negative Effects*

The academic gamification research does not ignore game design elements causing negative effects. Algashami [39] cataloged various negative effects, which the author called "gamification risks", dividing them into five categories of risk factors: performance, societal & personal, goals, tasks, and gamification design. The author identified 20 gamification risks, amongst them: performance mis-judgments, lowering self-esteem, counterproductive comparison, lack of group coherence, lack of engagement, reduce the quality, and kill the joy. In comparison with our work, Algashami's research [39] is not focused on gamification elements and neither on gamification applied to education, but on risk factors' identification and management strategies in large-scale businesses using gamification in their workplaces.

Hyrynsalmi et al. [40] pointed a lack of secondary studies about the negative effects of gamification. They categorized adverse gamification implications into limiting and harmful issues: the first category discusses gamification limiting the full capabilities of an artifact, and the second concerns the harmful consequences of gamification. We also found examples of both categories in [41]. In the mentioned context of a lack of secondary studies on the harmful effects of gamification, we decided to focus our research on them.

Hence, we searched for related secondary studies (synthesis of their findings in Table 2) but noticed variations concerning our purpose. We found significant differences for at least one of the following: subject, data analysis, date range, or a lack of focus on the negative effects of game design elements in gamification. Peixoto and Silva's review [42] had a different focus, aiming at building a gamification requirements catalog connecting game design elements to Bartle's Personality Types. Majuri et al. [43] present a review of 128 empirical research papers on gamification of education and learning and point out an excessive focus on quantifiable performance metrics and positive aspects. However, their work is not focused on negative effects and only covers the literature until 2015. Klock et al. [32] did also not focus on negative effects, besides having a data range limited from 2013 to 2016. Alhammad and Moreno's secondary study [44] had its scope scoped to gamification in software engineering education.

Finally, the secondary study by Toda et al. [27] is the work closest to ours, as they also focus on negative effects. However, as we noticed a significant amount of work in recent years, and their study identified only 17 papers within the date range from 2012 to the first half of 2016. Considering the existence of new evidence, we identified the need for an update [45]. Nevertheless, we decided to run a completely new mapping study, to allow us to apply a search strategy following the advice provided by [15] and to address our specific purpose more precisely, e.g., focusing directly on game design elements and identifying the type of empirical studies that revealed the negative effects.

Indeed, comparing our mapping study to the previously conducted secondary studies, as can be seen in Table 3, we expanded the time range covered until the end of 2020 and our search strategy allowed us to significantly increase the number of papers covered (87) when compared to the number found (17) by the closest related paper about negative effects of gamification in education [27]. Details on our mapping study follow.

## 3. Systematic Mapping

According to Kitchenham and Charters [46], a systematic mapping is "a broad review of primary studies in a specific topic area that aims to identify what evidence is available on the topic". We follow the procedures and guidelines described in [46] and [47], using a hybrid search strategy [15], combining database search with backward and forward snowballing.

### *3.1. The Research Questions*

Our goal was to organize evidence regarding the negative effects of Game Design Elements (GDEs) in the context of gamified education/learning software. Therefore, we derived the following research questions:

Table 2: Synthesis of findings about negative effects found by related secondary studies

| **Authors and Title** | What It Found Regarding Negative Effects Of Gamification? |
|---|---|
| Peixoto and Silva's "A gamification requirements catalog for educational software: Results from a systematic literature review and a survey with experts"[42] | No directly related results – instead, they created a catalog of gamification requirements (dynamics, mechanics and components – in other words, GDEs) for educational software, and how relevant those requirements were for players classified according to Bartle's taxonomy. |
| Majuri et al.'s "Gamification of education and learning: A review of empirical literature"[43] | Made some reference to negative effects on qualitative papers. Regarding quantitative papers, it also found that 23 related null or equally positive and negative effects, and 3 were mainly negatively oriented. |
| Klock et al.'s"Does gamification matter? A systematic mapping about the evaluation of gamification in educational environments"[32] | Found one negative non-significant influence through grades, motivation, and satisfaction and two negative effects involving grades. |
| Alhammad and Moreno's "Gamification in software engineering education: A systematic mapping"[44] | Found two negative effects on student knowledge/performance and one negative effect on student engagement. They also identified four student engagement, one student knowledge/performance, and one socialization cases where there wasn't any gamification effect (one of those was considered without any effect thanks to a mix of positive and negative effects). |
| Toda et al.'s "The dark side of gamification: An overview of negative effects of gamification in education"[27] | Identified and mapped 4 negative effects through 17 mapped papers, involved with 12 GDEs. Effects were indifference, loss of performance, undesired Behavior, and declining effects (with varying related GDEs). |

RQ1 - *What game design elements cause which negative effects in the field of digital education/learning?*

RQ2 – *In what fields of digital education/learning were the negative effects of game design elements found?*

RQ3 – *Which types of empirical studies were conducted to assess the negative effects?*

Table 3: Comparison between the identified secondary studies and our study

| Paper's Title | Time Period Covered by The Mapping | Focused on Negative Effects? | Focused on Educational Software? | Papers found |
|---|---|---|---|---|
| Gamification of education and learning: A review of empirical literature [43] | Up to June/2015 | No | No | 26 |
| Does gamification matter? A systematic mapping about the evaluation of gamification in educational environments [32] | 2010 to November/2015 | No | Unclear | 14 |
| Gamification in software engineering education: A systematic mapping [44] | 2011 to June/2017 | No | Unclear | 9 |
| The dark side of gamification: An overview of negative effects of gamification in education [27] | 2009 to June/2016 | Yes | No | 17 |
| Our study | Up to the end of 2020 | Yes | Yes | 87 |

To properly answer RQ1, we divided it into three more focused questions, organizing information on the GDEs causing negative effects (RQ1.A) and which negative effects affect which kind of user (RQ1.B and RQ1.C). Finally, we answer RQ1 by mapping the GDEs against their reported negative effects.

RQ1.A – *What game design elements caused negative effects in the field of digital education/learning?*

RQ1.B - *What negative effects of game design elements were found affecting those interacting with the software as users or being in the role of a student?*

RQ1.C – *What negative effects of game design elements were found affecting those maintaining the software or being in the role of a teacher?*

### 3.2. Search Strategy

We decided to use a hybrid search strategy, combining a database search on Scopus with forward and backward snowballing [15]. Hybrid strategies were found to be capable of achieving an appropriate balance of precision and recall when looking for primary studies [15]. To design the search string for the database search on Scopus, we used the PICOC (Population, Intervention, Comparison, Outcome, Context) criteria [48] as follows:

- *Population:* gamification software.
- *Intervention:* game design elements.
- *Comparison:* none.
- *Outcomes:* negative effects.
- *Context:* education/learning.

After that, we extracted the basic terms from the PICO criteria (gamification, education/learning, negative effects) and added synonyms and related terms. We decided not to include the intervention's terms, as we conducted the database search based on title, keywords, and abstract, where details on game design elements could have been omitted.

We added the following synonyms and related terms:

- Gamification: *gamify, gamiFIed, gamifying*.
- Education/learning: *information, teaching, curriculum, pedagogy, didactics, training, instruction*.
- Negative: *damaging, prejudicious, prejudicial, detrimental, counterproductive, inappropriate, harmful, perilous, limiting*.

Finally, we applied AND and OR logic operators to connect the terms, resulting in the following search string:

*(gamification OR gamify OR gamified OR gamifying) AND (education OR learning OR information OR teaching OR curriculum OR pedagogy OR didactics OR training OR instruction) AND (negative OR damaging OR prejudicious OR prejudicial OR detrimental OR counterproductive OR inappropriate OR harmful OR perilous OR limiting)*

As snowballing support tool, we used Publish or Perish [49], a software program that allows retrieving academic citations using information from Scopus and Google Scholar.

### 3.3. Inclusion and Exclusion Criteria

The inclusion and exclusion criteria can be found in Table 4 and 5, respectively. The exclusion criteria were derived from the inclusion criteria and provide details on our three-phase filtering procedure. To organize and filter the documents throughout the systematic mapping, we used Rayyan, a free web application to support systematic review authors [50].

Table 4: Inclusion Criteria

| **Inclusion Criteria** | **Reasoning** |
|---|---|
| IC1 - Papers which include negative effects of GDE applied in the field of education/learning in the context of gamification | *Research subject* |
| IC2 - Papers which passed through peer review | *To ensure a minimum level of quality* |
| IC3 - Papers in English | *Quality verifiable by the authors of this study* |

### 3.4. Applying the Search Strategy

#### 3.4.1. Search strategy application reported in our previous study

We first applied the search string on Scopus on July 28th, 2020, searching within the title, abstract, and keywords. It returned 180 documents, upon which we applied the exclusion criteria through three filtering phases, as described in Table 6. After this initial filtering a set of 64 papers remained.

Thereafter, still as part of our conference paper efforts [14], we conducted backward and forward snowballing using these 64 papers as seed set, both on

Table 5: Exclusion Criteria of the Three Filter Phases

| **Exclusion Criteria** | **Filter Phase** | **Reasoning** |
|---|---|---|
| EC0 – Papers not in English | First Filter Phase | *Quality not verifiable by the authors of this study* |
| EC1 - Paper which were not about effects of GDE applied in the field of education/learning | First Filter Phase | *Not about the research subject* |
| EC2 - Duplicated papers | First Filter Phase | *Duplicated* |
| EC3 - Papers that did not report negative effects | Second Filter Phase | *Not about the research subject* |
| EC4 - The paper has a more up to date version (e.g., journal extension) | Second Filter Phase | *Between two peer-reviewed versions reporting the same results, the most recent is to be used* |
| EC5 - The paper is grey literature | Second Filter Phase | *Typically not peer reviewed* |
| EC6 - The paper represents a secondary or tertiary study | Third Filter Phase | *Our study is a secondary study* |
| EC7 - The paper is mainly about the non-digital use of GDE | Third Filter Phase | *Focus of this paper is on digital artifacts* |
| EC8 - The paper is a short paper (less than 4 pages) | Third Filter Phase | *Typically does not represent complete research results* |
| EC9 - The paper was inaccessible to the authors | Third Filter Phase | *No means to access the paper* |
| EC10 - Books and chapters are off | Third Filter Phase | *Problems with verifying the quality* |

August 18th, 2020. The papers retrieved from backward snowballing and from forward snowballing using Scopus citation information were merged with the seed set, resulting in 2338 unique entries. Additionally, considering that Mourão *et al.* [15] suggest using Google Scholar for forward snowballing, besides doing it using Scopus citation information, we also conducted forward snowballing using citation information from Google Scholar (on September 4th, 2020). The forward snowballing through Google Scholar found 738 additional unique entries. Hence, we ended up with 3076 unique entries (including the seed set 64).

We applied our inclusion and exclusion criteria to the title, abstract, and

Table 6: Filtering after Scopus database search based on title, abstract, and keywords [14].

| Scopus database search | *180* |
|---|---|
| **Removed because of** | **Amount** |
| EC0 | *3* |
| EC1 | *88* |
| EC3 | *6* |
| EC5 | *5* |
| EC6 | *9* |
| EC7 | *1* |
| EC8 | *1* |
| EC10 | *3* |
| Remnants of the Initial Search Phase and the Filter Phases | 64 |

keywords of the 3012 papers retrieved through snowballing, as shown in Table 7.

Table 7: Filtering of 3076 unique entries retrieved from snowballing [14].

| Removed because of | Amount |
|---|---|
| EC0 | *113* |
| EC1 | *2192* |
| EC2 | *83* |
| EC3 | *28* |
| EC4 | *1* |
| EC5 | *27* |
| EC6 | *177* |
| EC7 | *1* |
| EC8 | *6* |
| EC9 | *20* |
| EC10 | *288* |
| **Papers to read (including the seed set of 64 papers)** | 140 |

After the title, abstract, and keyword filtering, we conducted full-text-based filtering for the remaining 140 papers. The result of this full-text-based filtering is shown in Table 8, resulting in a set of 68 included papers. Out of those, 32 were found by the initial Scopus search, 18 by forward snowballing, 15 by backward snowballing, and 3 were retrieved by both forward and backward snowballing. These numbers also help to illustrate how snowballing can be complementary to database searches. We emphasize that we conducted the full-text-based assessment only after snowballing on purpose, as we thought that applying snowballing on some additional closely related papers would not be detrimental. Nevertheless, this decision indeed increased our snowballing effort.

Table 8: Full-text-based filtering of the 140 papers [14].

| **Papers to read** | 140 |
|---|---|
| **Removed because of** | **Amount** |
| No access (even after requesting authors) | *6* |
| EC1 | *44* |
| EC3 | *13* |
| EC6 | *5* |
| EC6 | *1* |
| EC8 | *1* |
| EC10 | *2* |
| **Papers included** | 68 |

Finally, still as part of our previous effort reported in [14], to complement our search strategy, we compared our set of 68 included papers against the 17 papers included by [27]. While our set of 68 papers to be included comprised 29 papers ranging from 2012 to 2016, only seven of them were also included by [27]; i.e., their search strategy did not retrieve 22 papers reporting negative effects of gamification in education/learning software that were retrieved by our search strategy. On the other hand, our search strategy missed nine papers included in their mapping (the remaining one was retrieved but eliminated from our mapping for not being related to “digital” GDEs – EC7). As a result of this comparison, to present a mapping including all papers that we were aware of, we manually included the papers found by [27] that were missed by our search strategy, ending up with a final set of 77 included papers for our initial publication [14].

### *3.4.2. Extending the search efforts*

A natural extension strategy would be covering the gap of papers published until the end of 2020, and conducting additional forward and backward snowballing iterations.

As forward snowballing based on previously included papers is an effective strategy for updating systematic literature studies [51], the update until the end of 2020 could be accomplished by applying forward snowballing to the set of 68 papers identified initially as part of our strategy. Similarly, we could conduct additional backward snowballing iterations (i.e., on the papers retrieved through the first forward and backward iterations).

We checked upon the feasibility of applying this extension strategy with reasonable effort. We noticed that, a second snowballing iteration on the 68 papers, keeping our search temporally upper limited to the end of 2020, would involve analyzing a total of 5632 additional entries (1275 from backward snowballing and 4357 from forward snowballing, with a small overlap between both searches). Unfortunately, despite our best efforts from October 22, 2021 to January 18, 2022, this amount of entries proved to be unfeasible to handle as part of this extension. We understood that complementing snowballing iterations until saturation would involve analyzing several thousands of papers and characterize enough effort for a completely new paper. All the data from the unfinished second snowballing iteration is available in our Zenodo repository (www.doi.org/10.5281/zenodo.6279062). Therefore, we decided on a different strategy to assure feasibility within a reasonable manuscript extension effort.

With the intent to address the gap involving papers from the 2nd semester of 2020 with reasonable effort, our strategy involved: (i) re-executing our original search string on Scopus, limiting results until the end of 2020; (ii) filtering these papers; and (iii) applying forward and backward snowballing on the additional included papers.

Thus, in January 18, 2022 we re-executed our original search string on Scopus, searching within the title, abstract, and keywords. The search retrieved 266 documents. Out of these, 59 were excluded for being from 2021 or beyond. We identified that 176 of these papers were also identified in our similar search conducted in July 28, 2020 (we cannot explain why it did not retrieve all 180 previously returned papers, as we executed the exact same search string). The title, abstract, and keyword based filtering of the remaining ones is shown in Table 9.

Table 9: Closing the gap - Filtering after Scopus database search based on title, abstract, and keywords.

| **Scopus database search** | *266* |
|---|---|
| **Removed because of** | **Amount** |
| Papers from 2021 and beyond | *59* |
| Papers retrieved in July 28th, 2020 | *176* |
| EC1 | *7* |
| EC3 | *3* |
| EC6 | *3* |
| EC10 | *3* |
| Remnants of the Initial Search Phase and the Filter Phases | 15 |

The exclusion criteria applied was the same as before, with the difference that we did one filtering phase instead of three, covering all the exclusion criteria. After full-text-based filtering (as shown in Table 10) a set of 4 papers remained.

Table 10: Full text filtering of 15 unique entries

| **Removed because of** | **Amount** |
|---|---|
| EC1 | *9* |
| EC3 | *1* |
| EC10 | *1* |
| **Papers included** | 4 |

Then, we conducted backward and forward snowballing using these 4 papers as a seed set, on January 29, 2022. Google Scholar's backward snowballing retrieved 192 unique entries, Scopus forward snowballing retrieved 11 and Google Scholar forward snowballing retrieved 2. In total, backward and forward snowballing retrieving 205 entries.

We applied the title, abstract, and keyword filtering on these 205 entries, resulting in 44 papers, as shown in Table 11. Thereafter, we conducted full- text-based filtering for the remaining 44 papers, as shown in Table 12, resulting in a set of 7 additional papers. Including the 4 papers obtained from the search string filtering, the search strategy extension identified 11 additional papers. Extending the set of papers retrieved through our search from 68 to 79.

At all, considering the original search effort and this extension, out of the

Table 11: Closing the gap - Filtering 205 entries retrieved from snowballing.

| **Removed because of** | **Amount** |
|---|---|
| Papers from 2021 and beyond | *8* |
| EC0 | *14* |
| EC1 | *63* |
| EC2 | *20* |
| EC3 | *2* |
| EC4 | *3* |
| EC5 | *6* |
| EC6 | *10* |
| EC10 | *34* |
| **Remnants of the 1st Filter Phases** | *44* |

Table 12: Full text filtering of 44 entries

| **Removed because of** | **Amount** |
|---|---|
| EC1 | *22* |
| EC2 | *9* |
| EC9 | *2* |
| EC10 | *4* |
| **Papers included** | 7 |

79 papers, 36 papers were found by the initial Scopus searches, 18 by forward snowballing, 22 by backward snowballing, and 3 were retrieved by both forward and backward snowballing. Again, these numbers help to illustrate how snowballing can be complementary to database searches.

When comparing our new set of 79 included papers against the 17 papers included by [27], while our set of 79 papers to be included comprised 34 papers ranging from 2012 to 2016, only eight of them were also included by [27] i.e., their search strategy did not retrieve 26 papers reporting negative effects of gamification in education/learning software that were retrieved by our search strategy. On the other hand, our search strategy missed eight papers included in their mapping (the remaining one was retrieved but eliminated from our mapping for not being related to “digital” GDEs – EC7).

Indeed, after the second search that found the eleven additional papers, we found out that of the nine papers missed by the original search, one appeared as part of the backward snowballing of the second search. Furthermore, we noticed that four additional missed papers were part of the set of papers to be analyzed for the second snowballing iteration. While, considering the effort, as previously justified, we could not apply snowballing until saturation, we believe that the remaining four papers found by [27] and not by us would probably be retrieved as part of subsequent snowballing iterations. On the other hand, the effort of snowballing may be unfeasible within popular topics of research, such as gamification, leading to several thousands of papers to be analyzed.

Aiming at present a mapping including all papers that we were aware of, we manually included the eight papers found by [27] that were missed by our search strategy, ending up with a set of 87 included papers.

In summary, the scope of our search strategy (analyzing more than 3500 papers) and the added value (we extended the previously mapped evidence from 17 to 87 papers) provides an unbiased and meaningful overview on the adverse effects of gamification in educational software.

### *3.5. Data Extraction*

We extracted data from the 87 included papers focusing on answering our research questions. We used Google Sheets to organize the extracted data. To increase reliability in our extraction process, the spreadsheet with all the extracted data is available in an online Zenodo open science repository (www.doi.org/10.5281/zenodo.6279062).

We answered RQ1 by extracting data for RQ1.A, RQ1.B, and RQ1.C and connecting the GDEs with their respective negative effects. For RQ1.A, we extracted the GDEs that were related to negative effects. For RQ1.B and RQ1.C, we respectively extracted the negative effects caused to main users (in this case, students) and those maintaining the system working or in the position of teachers. We followed the open coding guidelines proposed in [52] to assign the text of the papers to design elements and negative effects. During this process, different terms perceived as related to the same element or effect were associated with a single code. During this process we were conservative, avoiding to group codes that could potentially refer to different concepts. In case of doubt concerning coding, discussions were held among three of the authors of the present work.

To answer RQ2, we extracted the fields of education/learning where the gamified software were used (e.g., computer science, medicine). Finally, to answer RQ3 we extracted the types of empirical studies conducted within each paper reporting the negative effects.

## 4. Systematic Mapping Results

Figure 1 shows the distribution of the 87 included papers throughout the publication years. Results for each of our research questions based on the extracted data follow.

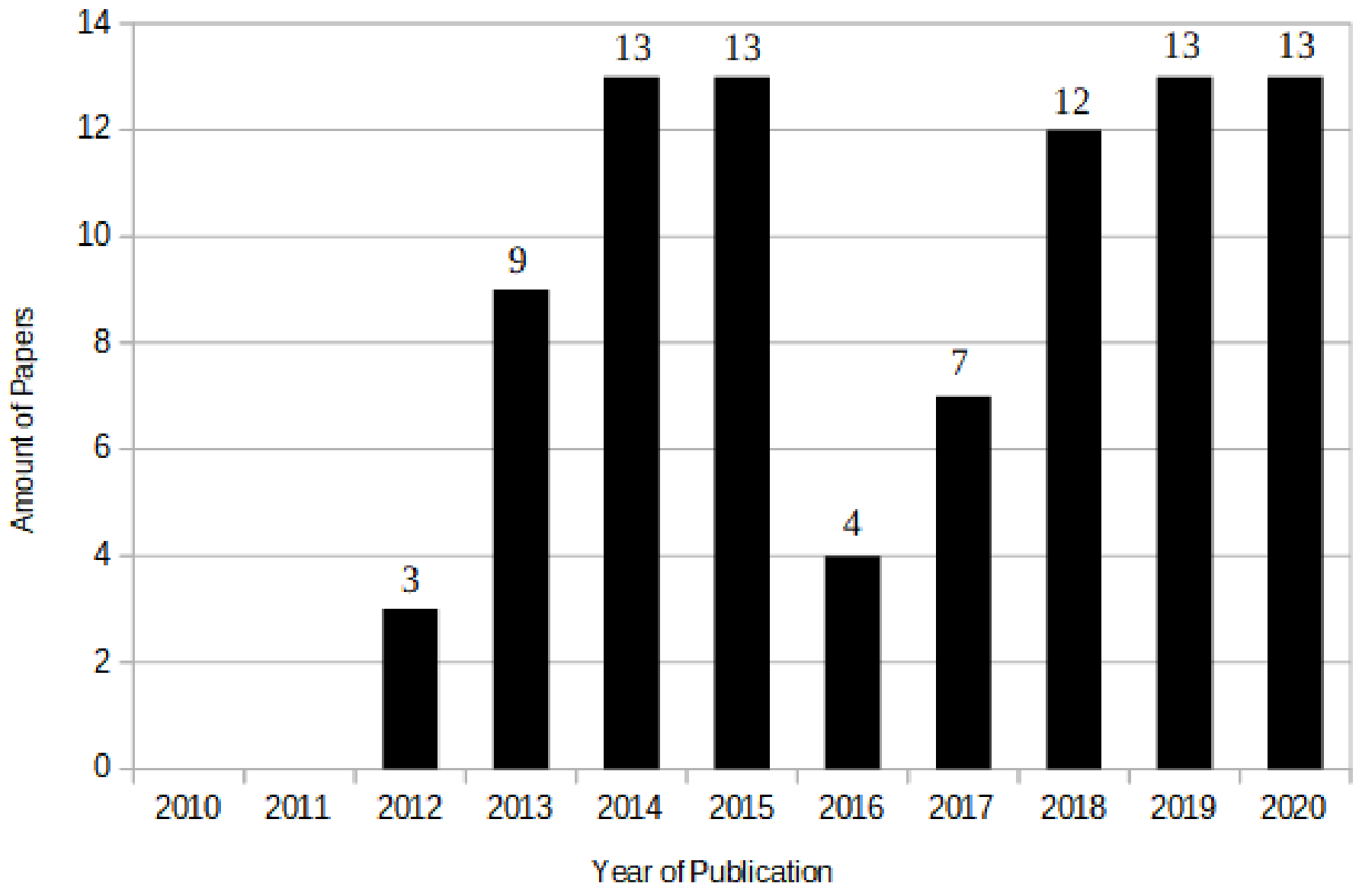

Figure 1: Publication years of the identified papers.

### 4.1. *RQ1 - What game design elements cause which negative effects in the field of education/learning?*

Overall, the papers reported 94 different GDEs, 69 different negative effects caused to the user, and ten different negative effects caused to the person maintaining the software or in the role of a teacher.

We decided to conservatively ground the answer to this question on associations reported by more than one paper instead of correlation and causation, strengthening, in the process, our confidence in the results. We adopted this approach, considering the vast amount of different GDEs and effects, plus the fact that many of the papers found had GDEs grouped and used together instead of individually, plus the fact that future research may show that some negative effects may be caused by a poor implementation of a GDE, instead of the GDE itself. The complete extracted data, allowing different analyses, can be found in our open science repository.

#### 4.1.1. *RQ1.A – What game design elements caused negative effects in the field of education/learning?*

Figure 2 summarizes the GDEs mentioned by at least two papers and the number of papers that referred to each of them as causing negative effects. The list of papers referring to each element can be identified in the online repository. It is possible to observe that most of the reported negative effects were associated with the use of badges, leaderboards, competitions and points. This makes sense given that these are GDEs commonly used in gamification, which may be related to creating competitive environments.

It is also noteworthy that there were several (77) other GDEs, which had only one paper each indicating negative effects. Further analysis is required to answer whether this can be explained by the lack of negative effects caused by these elements or the lack of investigations involving them.

#### 4.1.2. *RQ1.B - What negative effects of game design elements were found affecting those interacting with the software as users or being in the role of a student?*

Figure 3 shows the negative effects caused to the user mentioned by at least two papers and the number of times that papers referenced those negative effects. It is possible to observe that the most cited negative effects concern the lack of effect, worsened performance, demotivation, lack of motivation, and lack

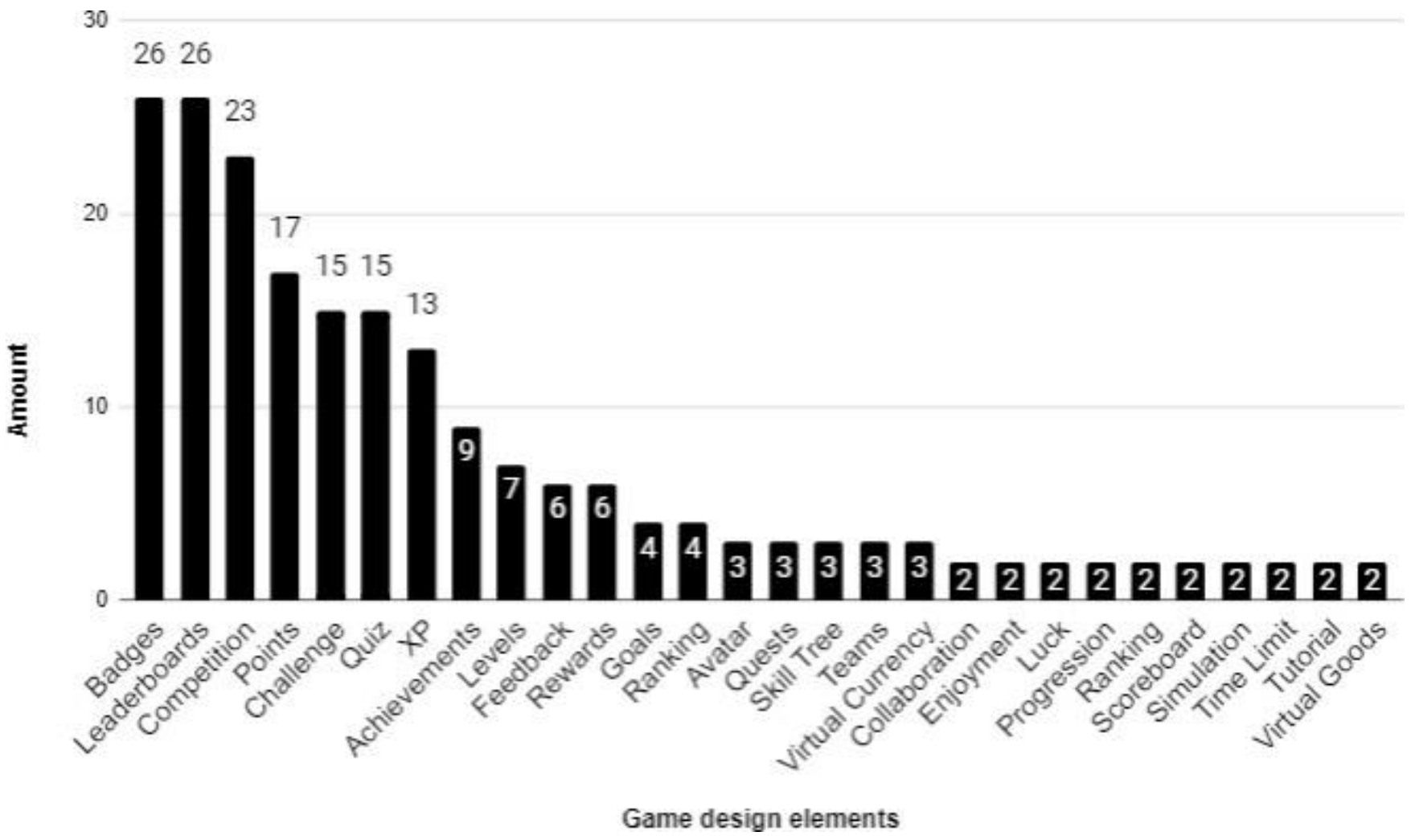

Figure 2: Game design elements and the amount of times they were reported being involved in negative effects.

of understanding.

The ethical issues of gaming the system and cheating were also recurrent reported effects, usually motivated by creating competitive reward environments (which stimulated users to break the rules to beat the competition) and/or systems with failures that enable users to easily score by cheating.

Discouragement and dislike of competition were also noticeable. Typically, if the student does not like competition and is losing in terms of grades, the visible gap between himself and those ahead will not result in improvement, but in losing hope [53].

Dislike of gamification and alienation or confusion for non-gamers appeared as well. Gamification is not a generic solution that works for any person, nor something that should be considered known by everyone. The ”alienation or confusion” cases made exactly that mistake and then had to deal with users that misunderstood how the system worked. For instance, a point-based system replacing a grading system, where the students did not understand how many points they needed to get a passing grade, and because of that lost many opportunities to increase their grades until it was too late [54].

Again, several (75) other negative effects caused to the user were mentioned only once.

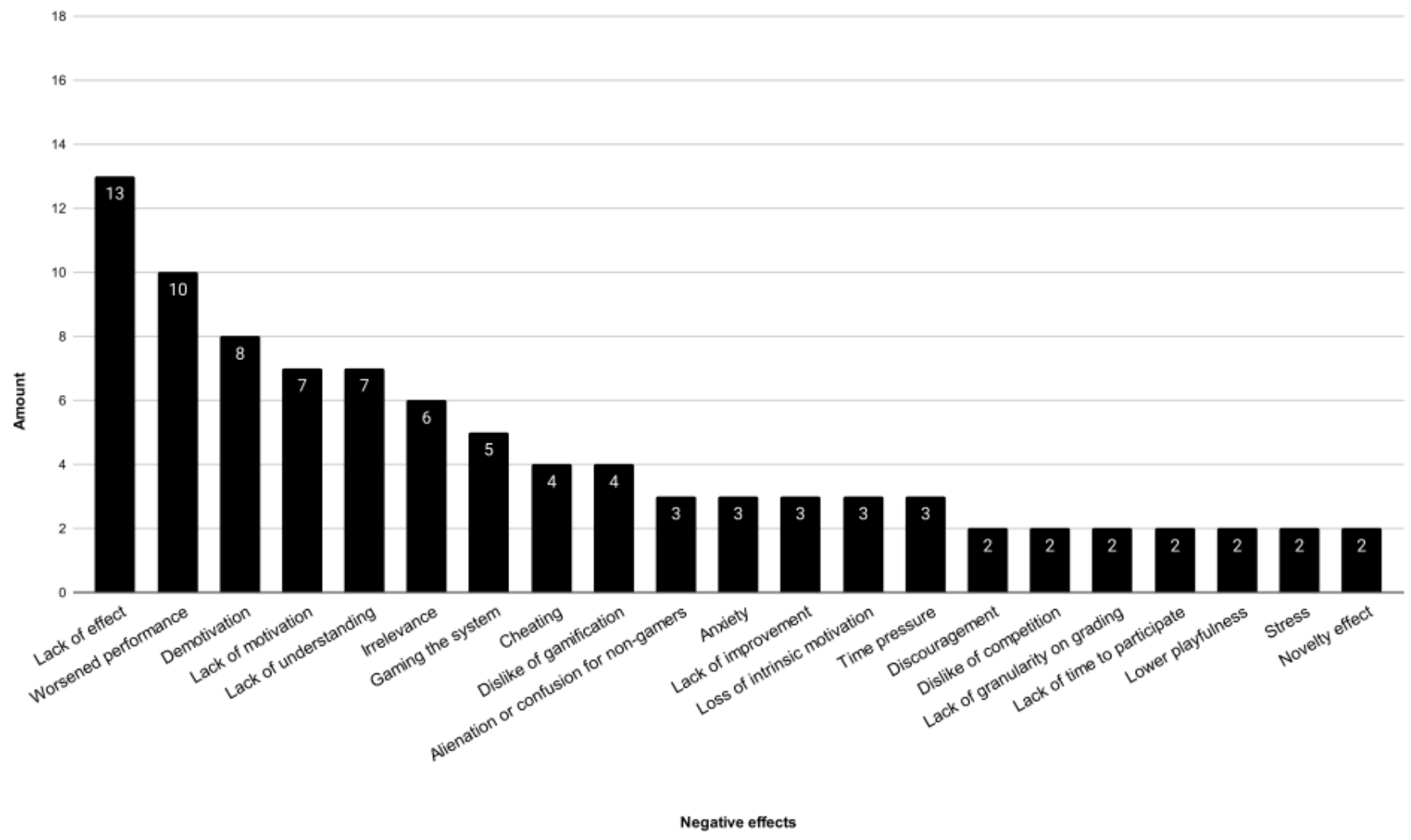

Figure 3: Amount of times negative effects reported being involved with game design elements, affecting those using the software as users or being in the role of a student.

Hence, the most common negative effect was that using the gamified software resulted in no difference when compared to not using the gamified software. Someone may argue that the negative effects characterized as being the “lack of” something are not negative, given that nothing bad effectively happened. However, for each of such neutral results to happen, gamification elements were designed and implemented, requiring human effort, that, while helping to show what does not work in terms of educational technology, effectively made no difference to the learners.

Another effect that needs explanation is the “Novelty effect”, which is a negative effect, in the sense that potential positive effects may be temporary. i.e., as soon as the user’s interest goes away, the positive effects will not apply anymore, and if they were not present for enough time, they may not be enough in terms of cost/benefit.

#### 4.1.3. *RQ1.C – Which negative effects of game design elements were found affecting those maintaining the software or being in the role of a teacher?*

In this question, we grouped both teachers and those who maintain the software working as both deal with the part of the gamified software that the students usually don't touch, and end up having to fix/circumvent problems that appear.

Table 13 shows the negative effects caused to those maintaining the software working or being in the position of a teacher, which were mentioned more than once and the number of times those negative effects were mentioned within the analyzed papers. It is possible to observe that the most common negative effects concern technical challenges and extra required effort or resources.

Table 13: Negative effects caused to the teacher/person maintaining the software working

| Negative effects caused to the teacher/person maintaining the software working | Amount |
|---|---|
| general technical challenges (bugs, difficulties with the software/hardware) | *7* |
| extra general human effort needed (e.g., money, time, people, effort) | *5* |
| engineering problems with the LMS (Learning Management System) used | *2* |
| Others | *7* |

Extra human effort and resources needed typically appear as a negative effect when the gamified software imply having to create additional content and taking care of additional tasks on top of the everyday tasks related to education. Finally, engineering problems typically appeared when learning management software did not cover what the designers wanted them to do, leading to implementation workarounds and potentially lower quality.[1] Seven other negative

[1]One of the "engineering problems with the LMS" was also counted as "extra general human effort needed", as it ended causing extra effort as well

effects were cited only once.

#### *4.1.4. Wrapping up RQ1 - What game design elements cause which negative effects in the field of education/learning?*

To complete the answer to RQ1, we mapped the GDEs against the related negative effects. The bubble plot in Figure 4 shows the GDE and negative effect pairings that appeared more than once in our systematic mapping. This mapping can help raise gamification designers' awareness of potential undesired negative effects of GDEs on education/learning software.

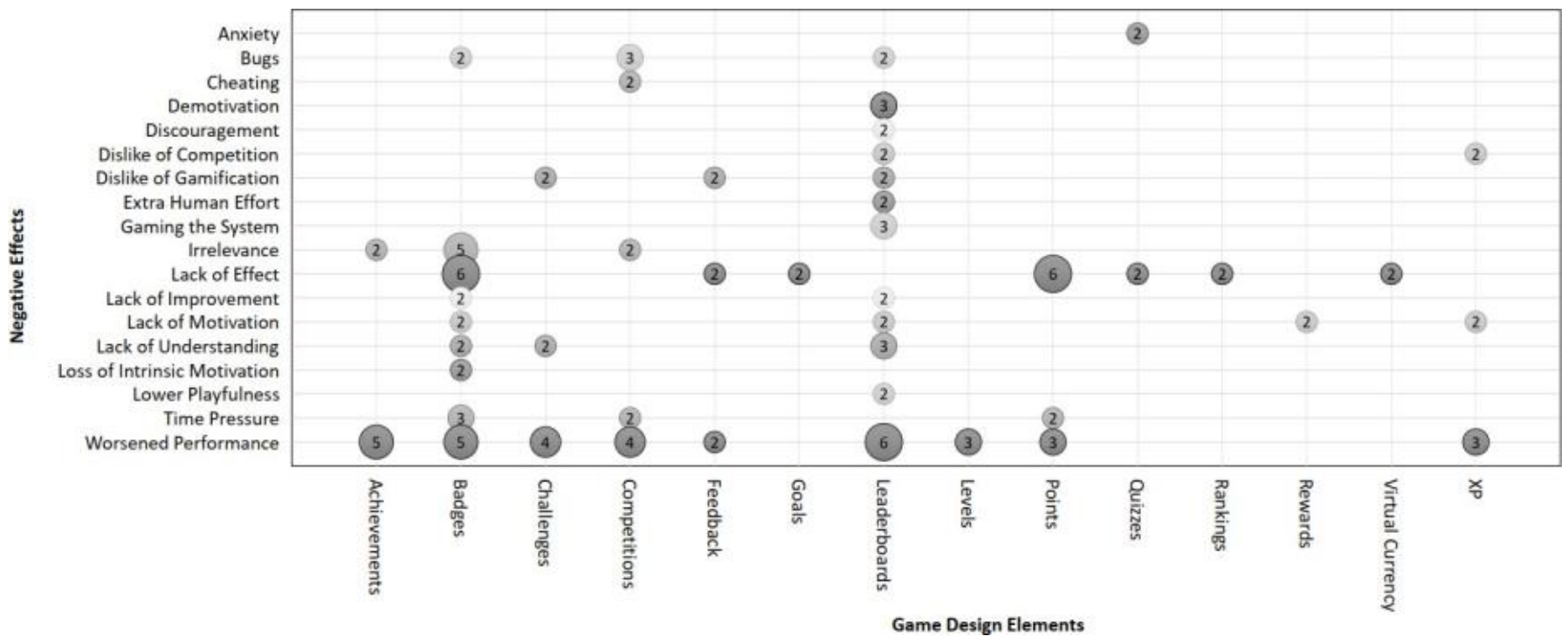

Figure 4: Bubble plot relating GDEs to negative effects.

It is possible to observe, for instance, that the use of Badges may have no effect [55][56][57][58][59][60], end up being irrelevant [55][61][62][63][64], or even result in worsened performance [59][65][66][67][68], amidst other reported negative effects, such as time pressure, bugs.

Similar interpretations can be made for the remaining GDEs. We provide a table containing the specific references related to the bubble plot of Figure 4 in Table 14. It is noteworthy that the primary studies included in our mapping vary in context and empirical strategy. Hence, a more in depth analyses for each GDE would involve carefully analyzing its related research papers, which are not easy to aggregate as part of a broadly scoped secondary study.

### *4.2. RQ2 - In what fields of education/learning were the negative effects of game design elements found?*

The fields where negative effects of GDEs were reported more than once are shown in Figure 5. Besides the listed ones, there were 29 other fields reported that were cited only once. It can be observed that the negative effects were reported in several different areas. Given the closeness between games, gamification, and digital technology, computer science being the most covered subject was expected.

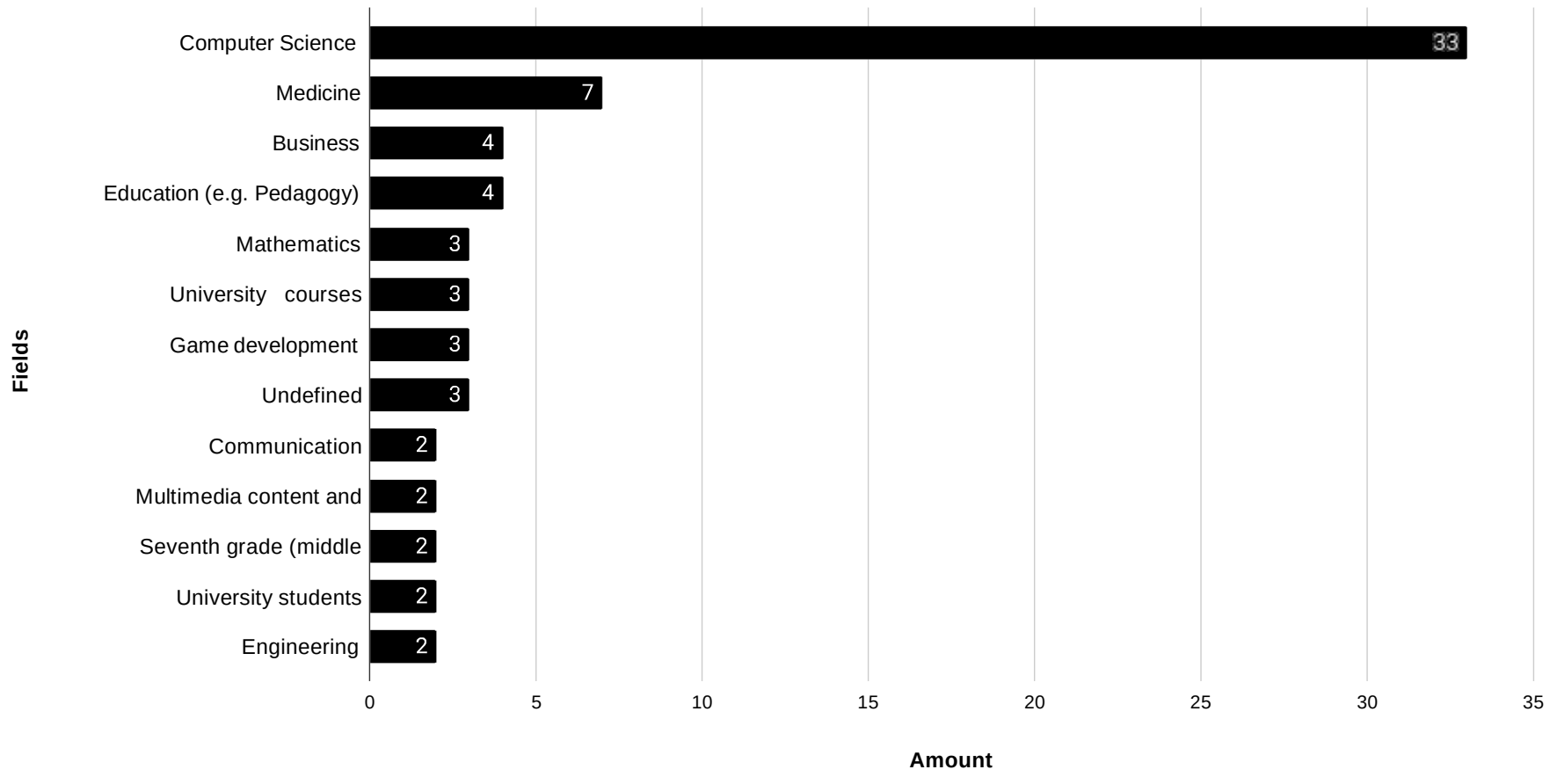

Figure 5: Fields of education/learning where negative effects related to game design elements were found.

### *4.3. RQ3 – What types of empirical studies were conducted to assess the negative effects?*

To answer this question, we used Wohlin's classification [105], which divides empirical strategies into surveys, case studies, and controlled experiments. As shown in Table 15, most of the research was reported as concerning case studies or controlled experiments, complemented by surveys. The positive aspect is that all papers reported applied at least one empirical strategy, which is expected in papers concerning the observation of effects.

Table 14: References of the bubble plot

| Game Design Elements | Negative Effect | Papers |
|---|---|---|
| Achievements | Irrelevance | [62] [69] |
| | Worsened Performance | [66] [69] [70] [67] [71] |
| Badges | Bugs | [55] [65] |
| | Irrelevance | [55] [61] [62] [63] [64] |
| | Lack of Effect | [55] [56] [57] [58] [59] [60] |
| | Lack Of Improvement | [72] [73] |
| | Lack of Motivation | [74] [75] |
| | Lack of Understanding | [59] [61] |
| | Loss of Intrinsic Motivation | [76] [77] |
| | Time Pressure | [61] [78] [79] |
| | Worsened Performance | [59] [65] [66] [67] [68] |
| Challenges | Dislike of Gamification | [70] [80] |
| | Lack of Understanding | [79] [81] |
| | Worsened Performance | [65] [66] [70] [80] |
| Competitions | Bugs | [65] [82] [83] |
| | Cheating | [81] [84] |
| | Irrelevance | [62] [64] |
| | Time Pressure | [78] [79] |
| | Worsened Performance | [65] [66] [70] [80] |
| Feedback | Dislike of Gamification | [70] [85] |
| | Lack of Effect | [57] [86] |
| | Worsened Performance | [65] [70] |
| Goals | Lack of Effect | [56] [87] |
| Leaderboards | Bugs | [65] [88] |
| | Demotivation | [65] [66] [54] |
| | Discouragement | [61] [81] |
| | Dislike of Competition | [75] [89] |
| | Dislike of Gamification | [70] [90] |
| | Extra Human Effort | [54] [91] |
| | Gaming the System | [72] [88] [92] |
| | Lack Of Improvement | [72] [73] |
| | Lack Of Motivation | [75] [93] |
| | Lack of Understanding | [59] [88] [92] |
| | Lower Playfulness | [94] [95] |
| | Worsened Performance | [59] [65] [70] [71] [68] [96] |
| Levels | Worsened Performance | [66] [68] [70] |
| Points | Lack of Effect | [57] [60] [86] [87] [97] [98] |
| | Time Pressure | [61] [78] |
| | Worsened Performance | [65] [66] [70] |
| Quizzes | Lack of Effect | [60] [99] |
| | Anxiety | [100] [101] |
| Rankings | Lack of Effect | [57] [102] |
| Rewards | Lack of Motivation | [103] [104] |
| Virtual Currency | Lack of Effect | [87] [97] |
| XP | Dislike of Competition | [75] [89] |
| | Lack of Motivation | [59] [75] |
| | Worsened Performance | [59] [68] [70] |

Table 15: Empirical Studies Conducted To Assess The Negative Effects (Counting Multi-Types As Different Entries)

| Types of empirical studies | Amount |
|---|---|
| Case Study & Survey | *37* |
| Controlled Experiment & Survey | *30* |
| Controlled Experiment | *9* |
| Case Study | *6* |
| Survey | *5* |

Overall, the mapping study identified empirical studies revealing negative effects of GDEs. It is possible to observe a concentration of research on a small subset of negative effects of educational software GDEs (*cf.* Figure 4 and Table 14). There is still limited research outside of that subset pointing to research gaps that the community could address.

Hereafter we enrich the discussion on the mapping study results through a focus group, debating the mapped negative effects of the GDEs with developers of gamified software.

## 5. Focus Group: Developer Perception on the Negative Effects of Game Design Elements

To complement our research with practitioner insights, we defined an additional research goal for this extended paper based on the Goal Question Metric template [106]: *Analyze* the mapping study results on negative effects of game design elements; *for the purpose of* characterization; *with respect to* the perception of software developers on the mapped negative effects of game design elements; and the perceived usefulness, ease of use, and intent of adoption of the mapping study results *from the viewpoint of* software developers of gamified software; *in the context of* their gamification experiences, gathered when developing a real gamified software involving several of the analyzed game design elements.

Based on this research goal, we defined the following research question: *What is the perception of software developers on the negative effects of using game design elements?* – By answering this research question, we want to characterize the developer's perception on the negative effect of game design elements reported in the literature.

For this purpose, we designed a focus group session for promoting in-depth discussion about the negative effect of game design elements. Focus group is a qualitative research method based on gathering data through the conduction of group interviews, called sessions, enabling the extracting experiences from the participants [107]. A focus group session is planned for addressing in-depth discussions about a particular topic during a controlled time slot. Additionally, focus group studies have been conducted in software engineering for revealing arguments and feedback from developers (e.g. [108, 109, 110]). Thus, we decided on using a focus group as a suitable option for understanding the developer's perception on the negative effect of game design elements.

### *5.1. Context and Participant Characterization*

We decided to conduct our focus group with experienced developers of the VazaZika development team [111, 112]. The VazaZika is a gamified software that encourages policies of education in public health concerning the prevention of mosquito-borne diseases such as Zika, Dengue, and Chikungunya. VazaZika aims to address the need for constantly reporting mosquito breeding sites. By using 12 game design elements (avatar, badge, challenge, comment, level, notification, point, ranking, social activity, social sharing, team and vote) and 16 game rules that reward citizens by reporting mosquito breeding sites the software successfully promote the collaborative work of citizens towards disease prevention, a fruitful competition among citizens, at the same time that improving the quality of reported mosquito breeding sites [111, 112].

The VazaZika software resulted from an international research project entitled *Leveraging Gamification and Social Networks for Improving Prevention and Control of Zika* [111, 112]. This project was performed by researchers in Software Engineering and Data Analytics from Brazil and the United Kingdom (UK). A total of 25 members participated in the project: one project manager; four development team leaders; 15 software developers, including the team lead-ers, distributed in two Brazilian cities, each with at least one developer per team; and seven senior researchers, five from Brazil and two from the UK. The project counted on the active contribution of a dozen Brazilian public health agents, which assisted many development activities.

From 15 software developers of the VazaZika development team, we successfully recruited four developers, including two team leaders. Table 16 provides a general view on the participants background. We have collected the table data from an online *Participant Characterization Form*, which we sent to participants minutes before the start of the focus group session. The team leaders are marked in Table 16 with the symbol (*).

Table 16: Participant Background Collected via Characterization Form

| **Question** | **P1*** | **P2*** | **P3** | **P4** |
|---|---|---|---|---|
| 1) What is your highest education level? | PhD | PhD | PhD | PhD |
| 2) How many years of experience do you have with software development? | 10 | 12 | 13 | 21 |
| 3) How many software projects have you worked on? | 6 | 20 | 6 | 10 |
| 4) How do you classify your level of knowledge with respect to Gamification? | High | High | Very low | Low |
| 5) How do you classify your level of knowledge with respect to Game Design Elements? | High | High | Very low | Low |

By observing Table 16, we can see that all participants have a high education level, all of them having a PhD in Computer Science. Also, all participants have at least ten years of experience with software development and have been involved with at least six software projects. Finally, two participants reported a high level of knowledge in gamification and game design elements (P1 and P2), and two others reported a very low (P3) and low (P4) level of knowledge. Thus, we assume that we have a balanced set of participants in terms of expertise on the specific topic. It is noteworthy that all of them actively participated as developers (or technical team lead - in the case of P1 and P2) in the gamification of the VazaZika software.

## 5.2. *Focus Group Design*

We carefully designed and performed our focus group by following the guideline proposed by Kontio et al. [107]. Figure 6 depicts the steps adopted throughout the focus group. We organized these steps in three major phases: (1) Preparation for the focus group session; (2) Discussion by group of game design elements; and (3) Discussion on the perception of the mapped negative effects. We describe each phase and step hereafter.

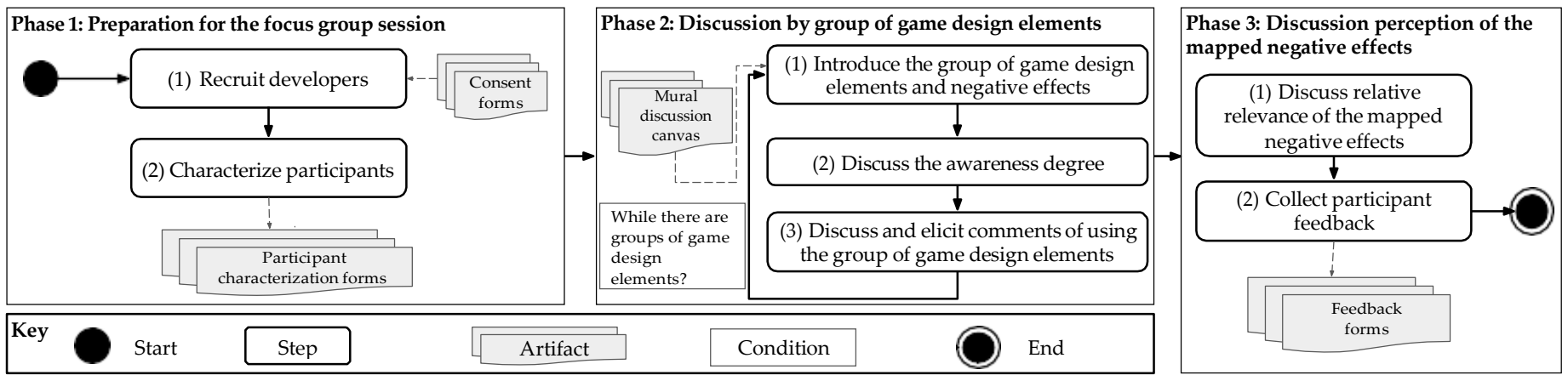

Figure 6: Overviews of the Focus Group Design.

**Phase 1: Preparation for the focus group session** – This phase consists of the collection of preliminary resources for supporting the execution of the focus group session. For this purpose, we follow two steps. **Step 1** consisted of *recruiting developers* with experience in gamification to engage in discussions. As mentioned in Section 5.1, we contacted developers of the VazaZika development team for participating in our study. We obtained the acceptance of four developers via *Consent Form* in which we explain our research goals, information about the waiver statements, and that the information provided by each participant will be treated confidentially and used for study purposes only. **Step 2** aimed to collecting basic information to *characterize the participants* via the *Participant Characterization Form*. Our major goal was profiling each developer, so we could better interpret our study results. We opted for asking short and simple questions in order to prevent participants from being tired or discouraged to participate in discussions right after filling out the form. More specifically, we collected data on the education level, years of experience with software development (academy and industry), the number of software projects they participated, and the level of knowledge about gamification and game design elements (see Table 16).

Due to the geographic distribution of our participants, we used an online environment to promote discussions on the negative effects of game design elements. Figure 7 illustrates the virtual template that we designed using the MURAL online tool[2]. In practice, by using this tool, we were able to build an interactive mural to facilitate the conduction of the focus group session. Our mural has eight well-defined sections. Sections A to G aimed at driving the discussion regarding the negative effects of game design elements.

In order to facilitate discussions, we have semantically grouped the game design elements by section as follow: (A) Badges and Rewards; (B) Competitions, Challenges, and Goals; (C) Leaderboards and Rankings; (D) Points, XPs, and Virtual Currencies; Feedback and Achievements; (E) Avatars; and (F) Quizzes. Finally, Section G aimed at driving the discussion and capturing the level of agreement on the *usefulness*, *ease to use*, and *intention to use* the information on potential negative effects of game design elements.

Figure 8 illustrates the template that we have defined for each aforementioned section. A section is composed of four parts (1, 2, 3, and 4). Part 1 contains a short description of the involved game design elements and the identified negative effects based on the literature. Part 2 is designed for capturing votes on the awareness degree concern the negative effects based on a four-point scale: *completely unaware*, *mostly unaware*, *mostly aware*, and *completely aware*. Finally, parts 3 and 4 were designed for participants to add notes on the *Pros* and *Cons* of using the game design elements, respectively.

**Phase 2: Discussion by game design elements** – This phase consists in collecting data regarding the developer's perception of the negative effects identified in the literature. As aforementioned, we have divide the game design elements in groups to facilitate the discussions and used a *Mural discussion canvas*. Thus, each group of game design elements is discussed in isolation. We defined three steps as follows. **Step 1** aimed at *introducing each group of game design elements, and the negative effects*. For this purpose, the moderator of the session read this information out loud (part 1 of Figure 8). Next, in **Step 2**, we asked each participant to assign one vote to the *awareness degree* of the group of game design elements according to the four-point scale depicted in part 2 of Figure 8. Each participant voted individually, without knowing the votes of his colleagues. At the end, the moderator showed the number of votes per awareness agree. Finally, in **Step 3**, we asked the participants to discuss and elicit *Pros* and *Cons* of using the game design elements. In this step, each comment should be documented as a note in the appropriate part: part 3 for positive comments and part 4 for negative ones, as shown in Figure 8. The note was just a brief summary of a *Pro* or a *Con*, typically taken within seconds, and we constantly asked participants to share knowledge and experiences surrounding the use of each game design element, to enrich the discussions and understandings. Additionally, whenever the moderator felt that a note is poorly written, he asked the participants to provide further considerations on the note.

[2]https://www.mural.co/

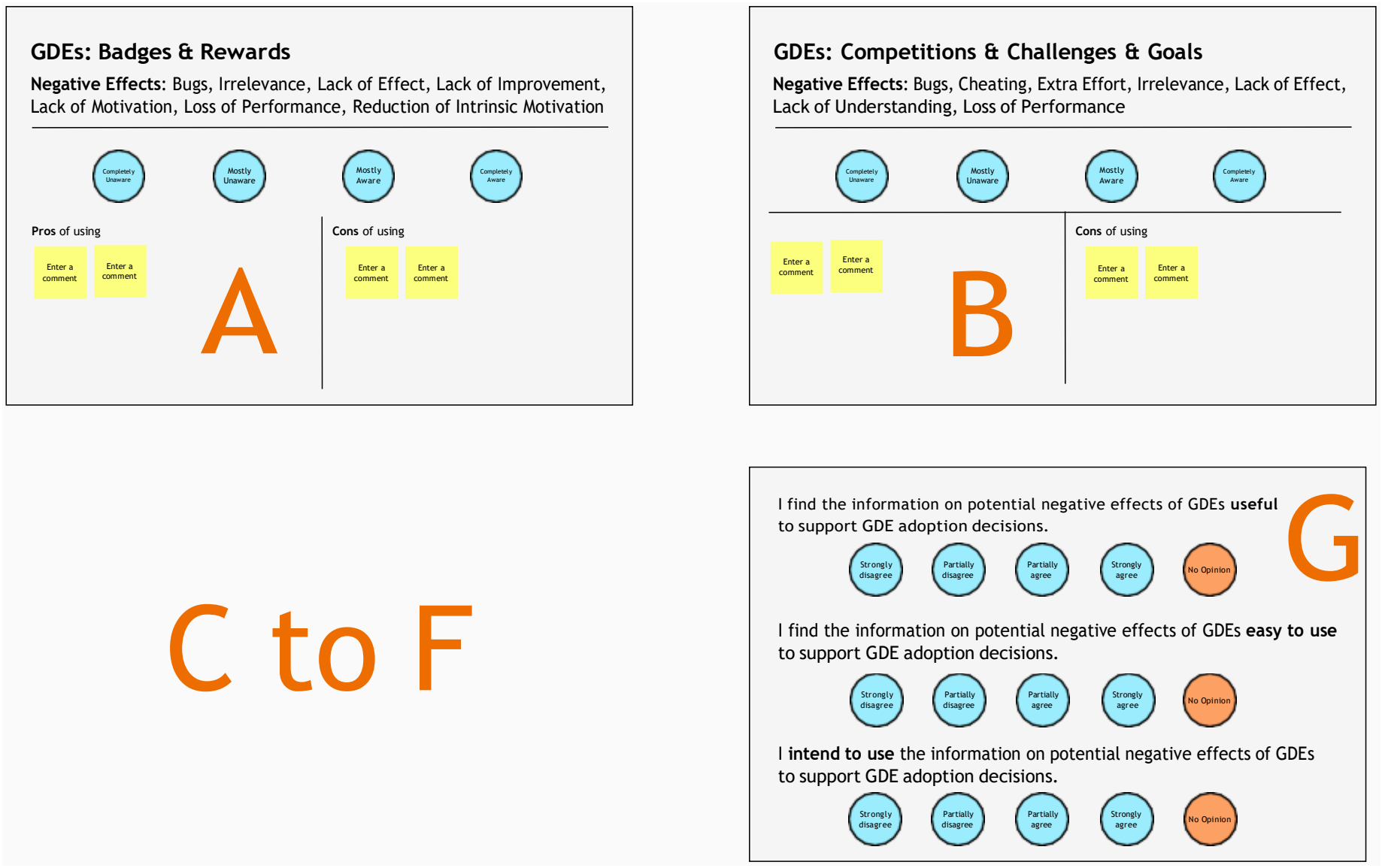

Figure 7: Focus Group Session Template Defined at MURAL.

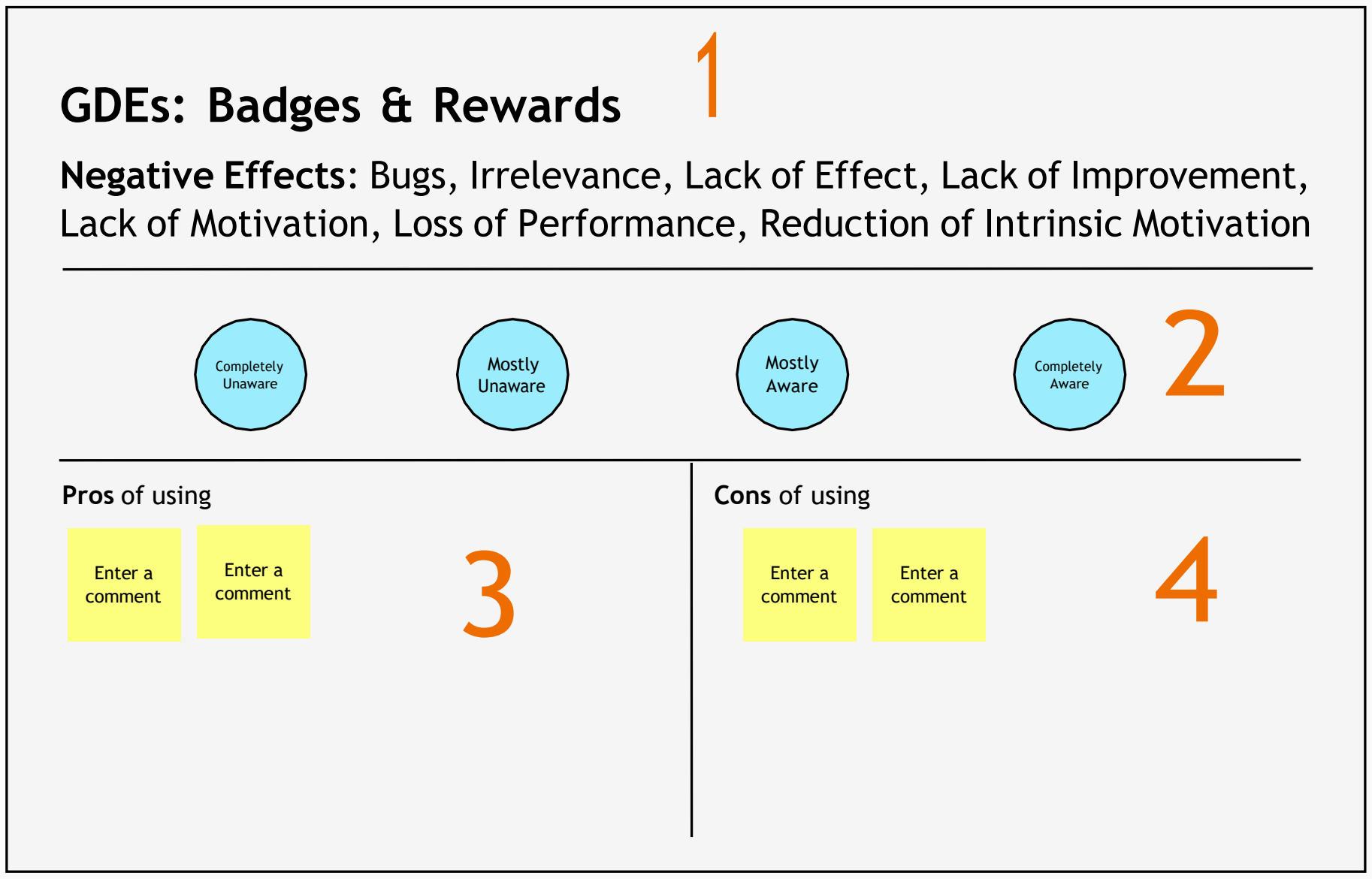

Figure 8: Template of each Section Defined at MURAL.

**Phase 3: Discussion about the perception of the mapped negative effects** – After discussing all groups of game design elements, the focus group session ended with a final discussion about the perceptions regarding the mapped negative effects. This phase has two steps described as follows. In **Step 1**, based on the main constructs of the Technology Acceptance Model (TAM) [113], we asked each participant to assign one vote to their agreement with each of the following statements (as illustrated in Section G of Figure 7): (1) I find the information on potential negative effects of game design elements *useful* to support game design element adoption decisions; (2) I find the information on potential negative effects of game design element *easy to use* to support game design element adoption decisions; and (3) I *intend to use* the information on potential negative effects of game design element to support game design element adoption decisions. We have used the five-point scale depicted in Figure 7 (G). Finally, **Step 2** aimed at collecting data about the participant's experience with the focus group session. Thus, by the end of the focus group session, we asked participants to fill out a *Follow-up Form*. We aimed at assuring that each developer felt confident and comfortable to discuss the negative effect of game design elements.

We emphasize that the focus group session was conducted online via a Zoom Meeting. Additionally, we kept video and audio records of the session to support the data analysis. We often accessed the video and audio records for understanding what developers meant with each note. The focus group session was conducted in January 24, 2022 lasting one and a half hours.

### *5.3. The Developers' Perception on The Negative Effects*

We asked participants to assign one vote to the *awareness degree* on the negative effects reported in the literature and to provide us with comments on pros and cons for using the game design elements under discussion. We have collected these comments through post-it notes added by the participants in the session's virtual mural (as illustrated in Figure 8). In order to analyze these comments, we first watched the video and audio records and transcribed them into plain text. Thereafter, we analyzed all post-it comments written by the participants and associated transcription quotes. In the following subsections, we summarize the comments that emerged for each group of game design elements.

#### *5.3.1. Badges and Rewards*

Figure 9 illustrates the comments made by the participants. In this group of game design elements, three participants reported being *mostly unaware*, and only one participant reported being *mostly aware* about the negative effects. We detail each comment as follows grouped by positive and negative ones.

*Positive comments.* The participants mentioned a total of two positive comments. **Increases Extrinsic Motivation.** Participant P1 mentioned that he was unaware about the negative effect of reducing intrinsic motivation. On the other hand, the participants mentioned that the use of these game design elements, in isolation or together with others, is important to increase extrinsic motivation. As mentioned by P1 and agred by P3, as follows.

*P1: "when we see that a player has many badges, what we expect as developers is that another player becomes more motivated, [...] the perception is that it increases*

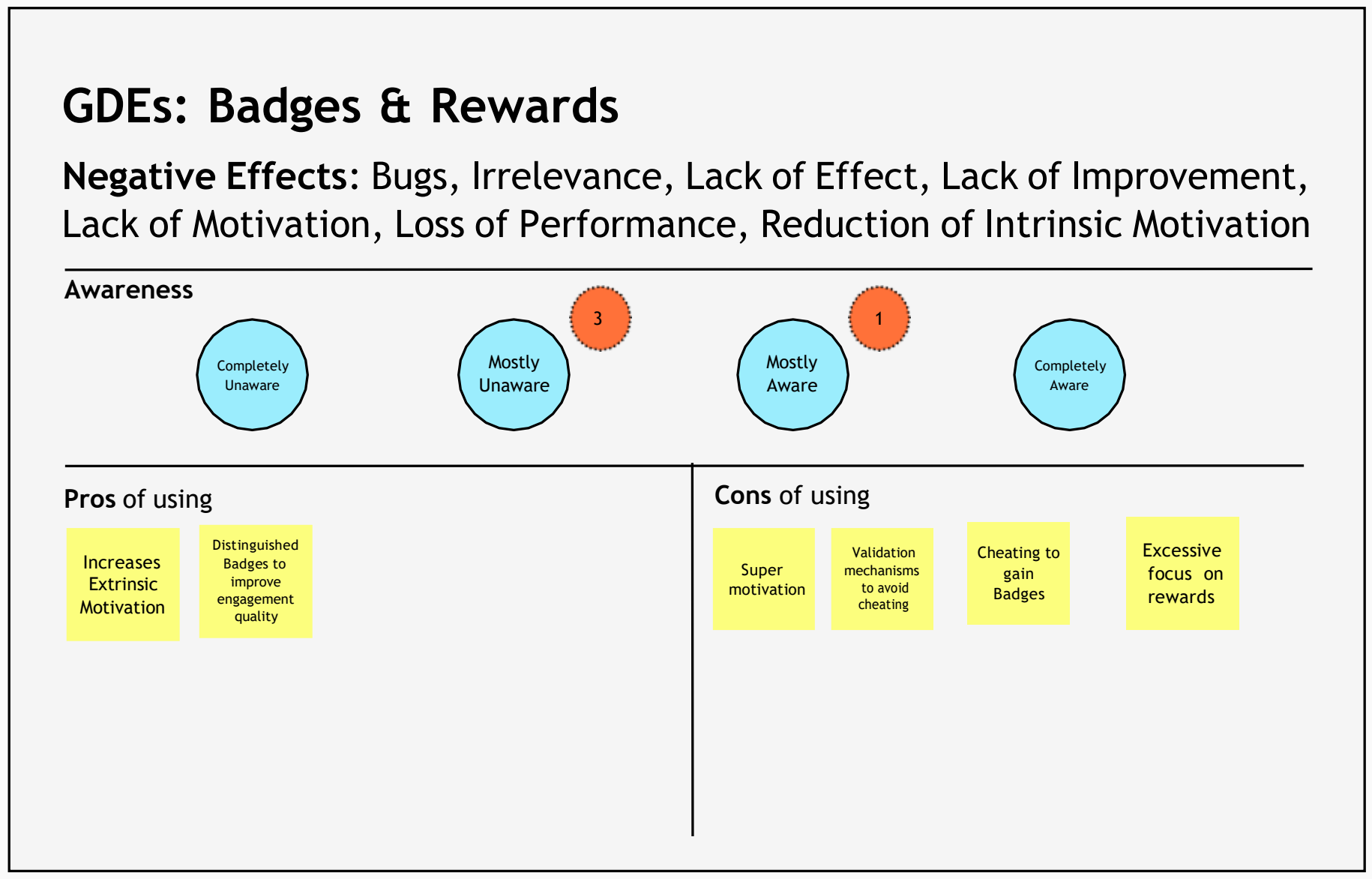

Figure 9: Positive and Negative Comments for Badges and Rewards.

*the extrinsic motivation. However, I was unaware that [the use of badges and rewards] could reduce intrinsic motivation."*

*P1: "We made some simulations using badges as part of the challenges, I would say that yes, it had a return on the motivation."*

**Distinguished Badges to improve engagement quality.** Another positive point highlighted by the participants was the use of distinguished badges to improve engagement quality. In this sense, participant P3 mentioned that one of the strategies used to motivate users of the VazaZika software was the use of distinguished badges, which could only be won by the user under specific conditions, associated with challenges and time intervals.

*P3: "[...] the idea of distinguished badges instead of [normal] badges that you can get doing a task anytime, [is that] you have distinguished badges associated with a certain time period, a certain challenge. So the idea would be to motivate certain periods associated with each badge."*

*Negative comments.* The participants mentioned a total of four negative comments. **Super motivation.** An interesting negative point raised by participants P2 and P3 concerns the super motivation of users in performing tasks, aimed only at winning badges and rewards. In this case, P3 also mentioned that super motivating users can lead to the introduction of false data into the software. For instance, in gamified software where the user earns badges for each check-in performed.

*P3: "[...] In the VazaZika software, we had a badge that the user would only earn if he was in a certain location in a specific period of time [...]. So it was possible to increase the [data] quality, because we knew that the user who was completing this challenge was reporting real mosquito breeding sites."*

Although the presence of super motivated users has a negative effect, participant P2 also mentioned that this effect can be minimized through the definition of validation mechanisms.

*P2: "[..] in programming sites in which the user is super motivated, writing code and sending a lot of code every time. The control mechanism is good [because] it tests and verifies that the code is working. Thus, the super motivated user is not a problem in this case. It would only be a problem if the control mechanism was bad."*

**Validation mechanisms to avoid cheating**. Another negative point mentioned by the participants was the need to introduce validation mechanisms to avoid cheating. As mentioned by participants P1 and P4.

*P1: "In order to avoid false data, we used the vote (up and downvote) to verify and validate if that reported mosquito breeding site was really real or not."*

*P4: "[...] we had a worry that the award by itself, without any type of control mechanism, could be harmful. In our context, at the health area, that involved the work of health public agent, government actions [...], a greater care was needed to ensure the fidelity of information."*

**Cheating to Badges.** Another point mentioned, and strongly related to super motivation, and the presence of validation mechanisms is the practice of cheating to badges, intending to earn more rewards as mentioned by the participant P2 as follows.

*P2: "[...] the player was misrepresenting the app data because he was super motivated to win the badge. However, I do not think that this avoids the bigger problem that is the false data that are entering the software, this is the big context."*

**Excessive focus on rewards.** Finally, the participants mentioned that the use of badges and rewards can generate an excessive focus on earning rewards. According to participant P4, this may be due to the gamification software being designed to generate rewards for users after performing simple and repetitive tasks, generating a deficiency in the game's logic.

*P4: "So, maybe it is related to certain gamified software that have more focus in provide rewards than the game logic."*

*P4: "[So], if the idea is to [provide] fun, entertain and attract [the players], we need to be more careful in giving rewards for each performed task by the player. This cannot go unnoticed. We need game logic too."*

### *5.3.2. Competitions, Challenges and Goals*

Figure 10 illustrates the pros and cons mentioned by the participants for Competitions, Challenges and Goals. In this group of game design elements, two participants reported a *mostly unaware*, and two participants reported *mostly awareness* about the negative effects. We detail each comment as follows grouped by positive and negative comments.

*Positive comments.* The participants mentioned a total of three positive comments as follows. **Communitary (social) integration.** P4 mentioned that the use of competition, challenges, and goals can increase community integration. More particularity, the participant mentioned the use of these game design elements in a particular community, for instance, the citizens of a particular neighborhood can increase integration between those involved. As we can see below.

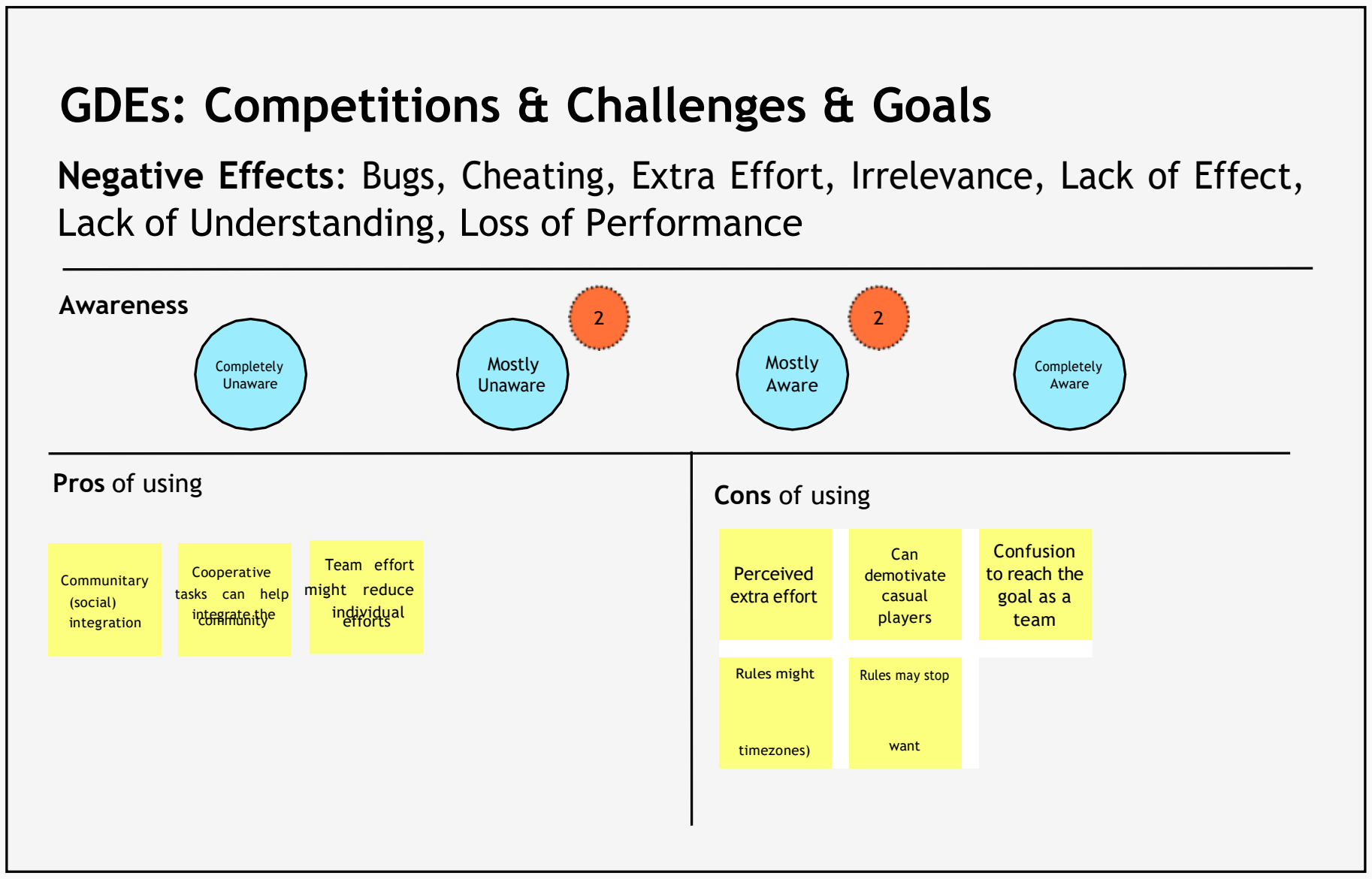

Figure 10: Positive and Negative Comments for Competitions, Challenges, and Goals.

*P4: "The context here influences a lot. I would add this type of competition as an advantage. So, if you are gamifying something that probably has a purpose of getting a result for a certain community, e.g., a industry, but specifically in our case citizens within a neighborhood or region, I see this as an opportunity for integration."*

**Cooperative tasks can help integrate the community.** In line with the previous positive reason, participant P2 mentioned that the use of cooperative competitions, i.e., in which players come together to achieve the same goal, is more motivating than just creating competitions between players.

*P2: "if you create competitions that aren't necessarily between players, but between players around a common goal, you can help players come together to achieve those goals. In VazaZika, we've done that by enabling team creation."*

**Team effort might reduce individual efforts.** Complementary, Participant P3 mentioned that depending on the type of challenge and its difficulty level, is essential that the challenge be done with two to three people. In this sense, the creation of cooperative tasks helps to reduce the effort to achieve a common goal.

> *P3: "What I think goes against the extra effort, is that depending on the challenge, you can not do it alone, you need two to three people"*

*Negative comments.* The participants mentioned a total of five negative comments, as follows. **Perceived extra effort.** Participant P1 has mentioned that extra effort is a negative effect when challenges require physical effort. In the context of the VazaZika software, citizens had to report and validate real cases of mosquito outbreaks in loco. The participant also mentioned that even though the challenge is done in teams, the extra effort was still perceived by the citizens.

> *P1: "Most of our challenges required a physical effort since the citizens had to go to a location to report or validate the existence of a mosquito breeding site, either individually or in teams. We were aware that there was an extra effort because they would have to move around."*

**Can demotivate casual players.** Another interesting negative reason mentioned by participant P2 concerns the demotivation of casual players to participate in competitions that attract frenetic or experienced players. In other words, a casual player may feel demotivated to compete with frenetic players, whose experience level is constantly increasing due to the time spent these players in performing gamified activities in the software.

> *P2: "You have a frenetic community of [players] who set the bar high, and for a [casual player] to participate in competitions is simply demotivating, because he cannot play casually in these competitions."*

**Rules might demotivate participation.** Another problem associated with the demotivation of casual players is related to the definition of rules that can restrict the participation of some players in collaborative challenges. For instance, rules that involve the availability of collaborative challenges in a given time interval may restrict the participation of players who are in different time-zones. As mentioned by Participant P3 as follows.

*P3: "Another problem is that there are some rules in competitions that make it difficult for players to participate, e.g., players who are in a completely different time zone and who want to compete with their friends from Brazil who are in another time zone. In this sense, a rule involving time restrictions discourages a player from participating in the competition."*

On the other hand, participant P4 mentioned that the definition of rules that restrict the participation of players is not always a bad design. That is the case of challenges that involve the participation of a community (a favela), or a small group of adolescents in a public school. Such challenges, although they have geographic restrictions, also stimulate the accomplishment of tasks or challenges pertinent to a certain location. For instance, a challenge to report mosquito outbreaks in a region with a public health policy deficiency.

*P4: "For a community geographically, a favela, adolescents from a public school [...] it would be a stimulus, to prevent diseases [...]"*

**Rules may stop competitors to participate in the way they want.** In line with the previous comment, participant P3 mentioned that some rules in addition to demotivating can stop players that want to participate in challenges. For instance, rules related to the minimum quality of participants for challenges. In this case, the software may be forcing challenges with certain characteristics that a player does not necessarily want as mentioned by participant P3 as follows.

*P3: "I remember that in the VazaZika software , some competitions could only be held by a team, and the team had a minimum amount of people. [...] So, the software*

*was forcing you to enter a competition with certain characteristics that you don't necessarily want."*

**Confusion to reach the goal as a team.** In fact the lack of understanding can be a negative effect. More specifically, participant P1 mentioned that although instructions for completing a challenge were explicitly provided to players, in some cases they could not understand what needed to be done to complete a challenge as a team.

*P1: "The lack of understanding could indeed have this negative effect. We tried to make it explicit what they had to do to complete a challenge, but they still couldn't understand."*

*5.3.3. Leaderboards and Rankings*

Figure 11 illustrates the pros and cons mentioned by the participants for Leaderboards and Rankings. Similar to the previous group of game design elements, two participants reported a *mostly unaware*, and two participants reported *mostly awareness* about the negative effects. We detail each comment as follows grouped by positive and negative comments.

*Positive comments.* The only positive comment mentioned was **Can create pro players**, by P2, where he argued that players with a higher level of ability in the game would be able to be beneficial outside it.

*P2: "Per example, trying to get the context of a gamified system, the HackerRank, (...) that is a programming questions' system, the pro players there... for you to be as good as them, you have to 'waste your life' writing code and submitting questions all the time, and being approved on the tests and gaining points."*
*P2: "But the pro players in those systems are so good in algorithms that they would easily be approved on certain business interviews, you understand? So, there's a lot of business monitoring those pro players, let's say that. So for the industry that is monitoring the rankings, this could be a positive point. Not considering all the negative effect of the thing that can stress you, stuff like that. For the corporate world this can help somehow."*

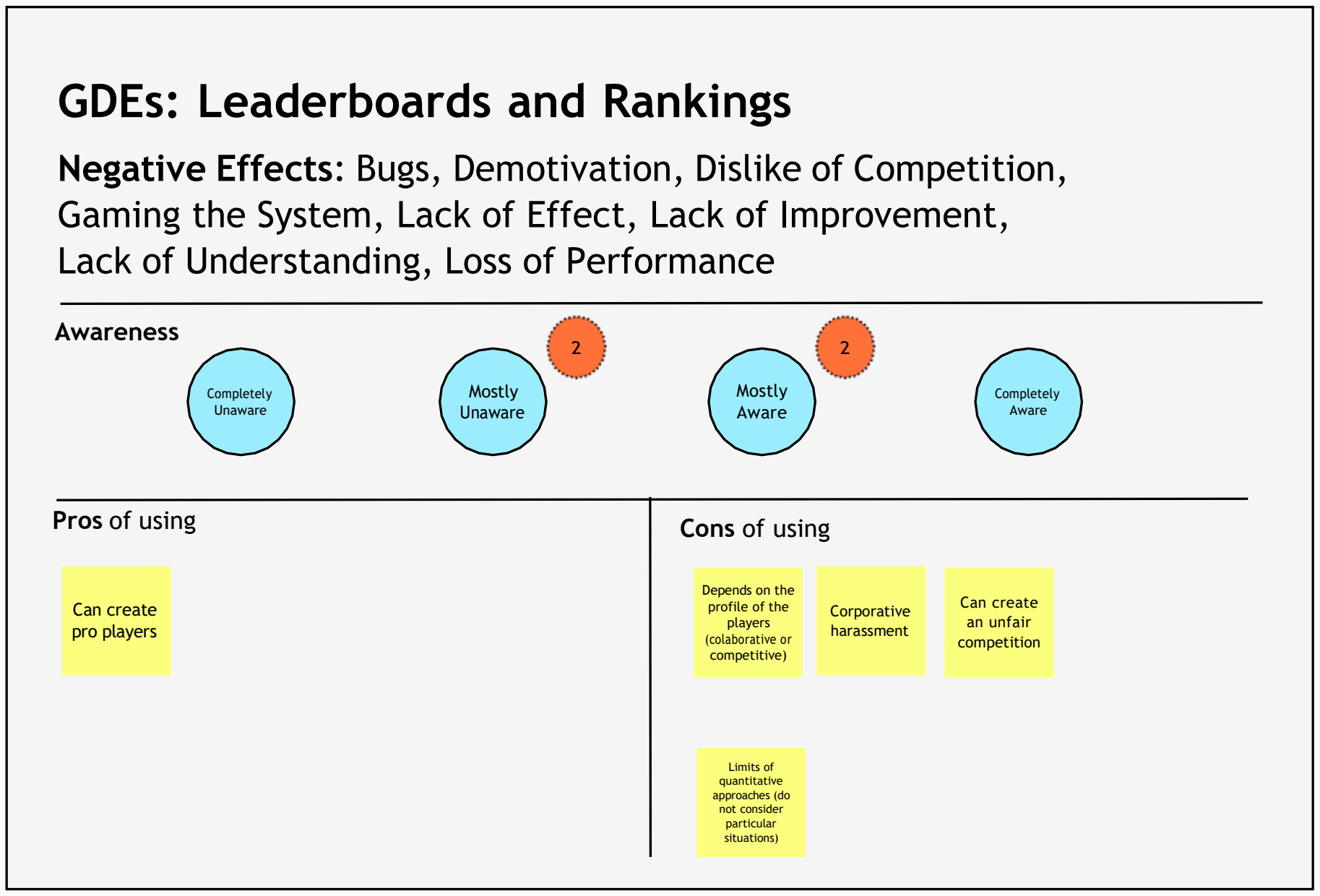

Figure 11: Positive and Negative Comments for Leaderboards and Rankings.

*Negative comments.* There are four negative comment, starting with **Depends on the profile of the players (collaborative or competitive)**, the participant P3 mentioned that depending on the type of player, the use of leaderboards or rankings can have a negative effect. For instance, unlike players with a competitive profile, who are motivated to get the best positions, collaborative players can feel unmotivated.

> *P3: "[...] there are those that will consider Leaderboards and Rankings as something that motivates them, "I want to be the first", but there are others with a more collaborative profile, their reaction will be "look, I don't want to participate and be the first", so it depends a lot of the [...] players' profile."*

**Cooperative harassment.** Participant P4 has mentioned that the use of leaderboards and rankings in a cooperative environment, mainly in the private sector, can lead to harassment or bullying of the player. In this case, participant P4 argued that in these environments, a player can use leaderboards and rankings as a form of harassment, both for those who are at the top of leaderboards/rankings and for those who are not.

> *P4: "[...] in the public sector and mixed economy, this worry with harassment, the person feeling excluded, diminished, or even demotivated. And that has a lot to do with the negative effect.[...] They see that as a way to reduce the other, both who is higher in the rankings/leaderboards and who is lower."*

**Limits of quantitative approaches (do not consider particular situations).** In line with the *Cooperative harassment*, participant P4 mentioned that a negative aspect is the use of leaderboards and rankings heavily based on quantitative approaches. P4 argued that in situations in which a player is unable to perform tasks in the software, he will lose positions in the ranking, which would be considered unfair. For instance, in cooperative environments when a player takes sick leave, his actions are no longer quantified.

> *P4: "[...]exceptions will always happen because we are human and there will be situations where the person will fall through the rankings and that will be considered unfair, as an example, a medic licence and the person went away. And the metrics are [...] quantitative, and the rankings are mostly quantitative."*

**Can create an unfair competition.** Is related to the demotivation of beginner players in achieving leadership or a similar position to players who are at the top of the leaderboard. In other words, depending on the beginner player's current level, he may never reach the leadership as mentioned by participant P1 as follows.

> *P1: "So I think this is a cons point, because depending on what level you are, you will never reach the lead."*

### *5.3.4. Points, XPs and Virtual Currencies*

Figure 12 illustrates the comments made by the participants for Points, XPs, and Virtual Currencies. About the awareness degree, three participants reported a *mostly unaware*, and one participant reported *mostly awareness* about the negative effects. We detail each comment as follows grouped by positive and negative comments.

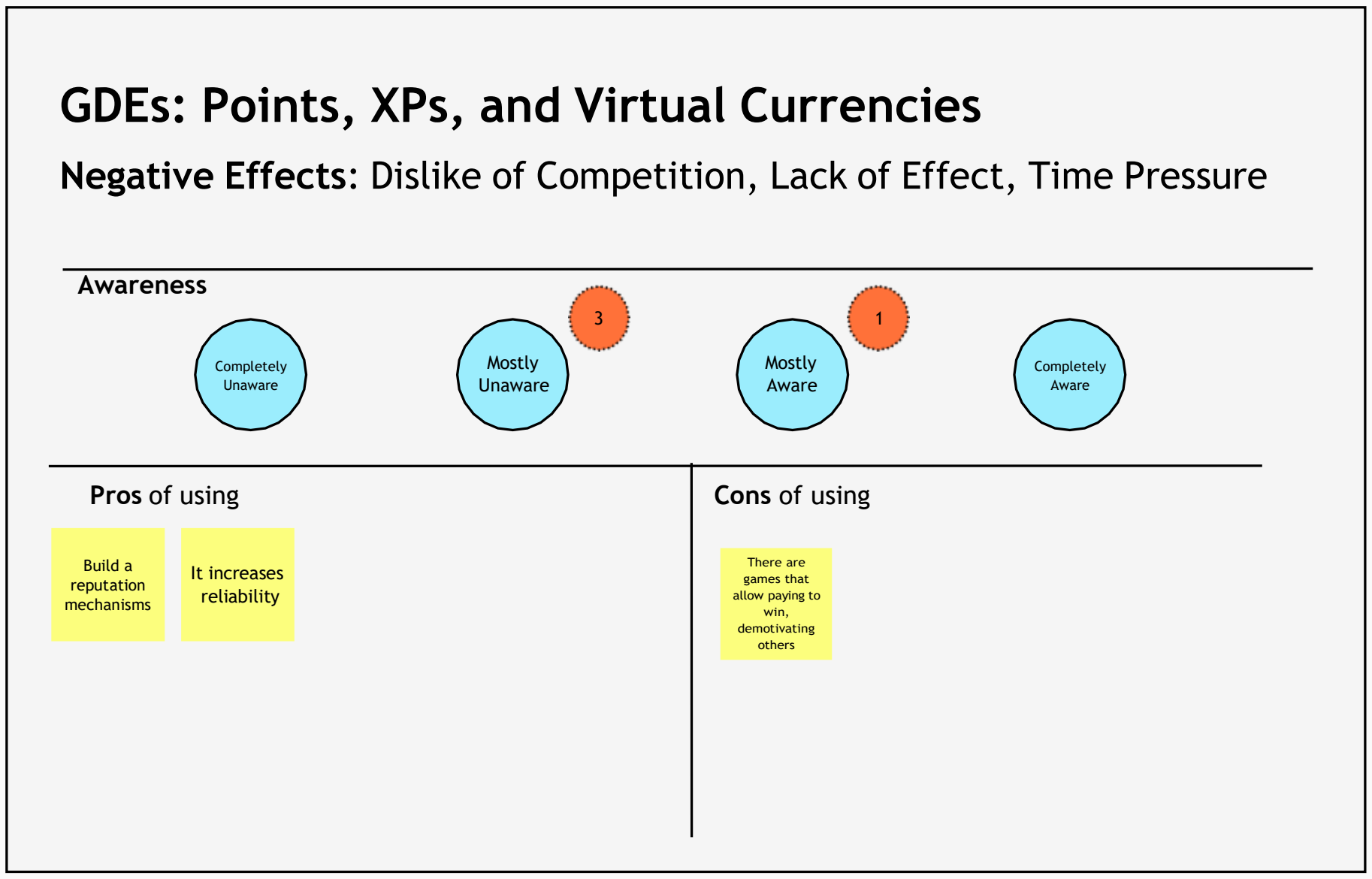

Figure 12: Positive and Negative Comments for Points, XPs and Virtual Currencies.

*Positive comments.* Two positive comments were made by participant P3: **Build a reputation mechanism** and **It increases reliability**. Participant P3 exemplified that in systems such as Stack Overflow, a gamified system of questions and answers about programming problems, the use of points helps to build a reputation in terms of the player's level of experience in response to specific answers. At the same time, responses from highly reputable players are considered highly reliable.

> *P3: "For instance, let's think about the case of Stack Overflow, that is a gamified system where sometimes you want to search an answer to a solution, then you have two possibilities, which one you choose first? The one of the user with the highest XP. [...]."*

*Negative comment.* The participants P2 and P4 mentioned that **There are games that allow paying to win, demotivating others** as a negative rea-son. In this sense, P2 discussed the negative effects of virtual currencies in allowing players to obtain advantages over other players (pay to win).

*P2: "It's a real option. And then you have games with a mechanic of paying to have advantages upon other players. This creates a demotivation. It's what people calls as pay to win."*

Participant P4 also argued that the use of virtual currencies, which facilitates the practice of pay to win, can discourage players who continually perform actions in the software in order to advance in the game.

*P4: "I agree a lot with P2. I can't see the gamification context, it does not occur to me that it would be worthy to give coin or the opportunity to pay to play, I can't see it, because usually the serious game, you will gamify a thing there to fix an issue, how are you going to add payment there? Why when you pay to achieve something without a fight for that, so that is done only to brag about the achievement. But in a serious game you have to do things, you have to solve problems. So, how can you in this context use a coin, not only as money but being able to pay to gain the money? Then in this pay to win, I agree completely with P2, and I can't see, can't glimpse in the serious game context that you would pay to win."*

### *5.3.5. Feedback and Achievements*

Figure 13 illustrates the comments made by the participants for Feedback and Achievements. In this group of game design elements, two participants were completely unaware, one participant was *mostly unaware*, and one participant as *mostly aware* of the negative effects. We detail each reason as follows grouped by positive and negative comments.

*Positive comments.* **Public feedbacks may serve as hints for other users** and **Achievements-driven players can play and contribute more**. P1 added that public feedbacks would be advantageous to engage the users. In other words, users can feel motivated, realizing that other users are performing tasks and being rewarded for them.

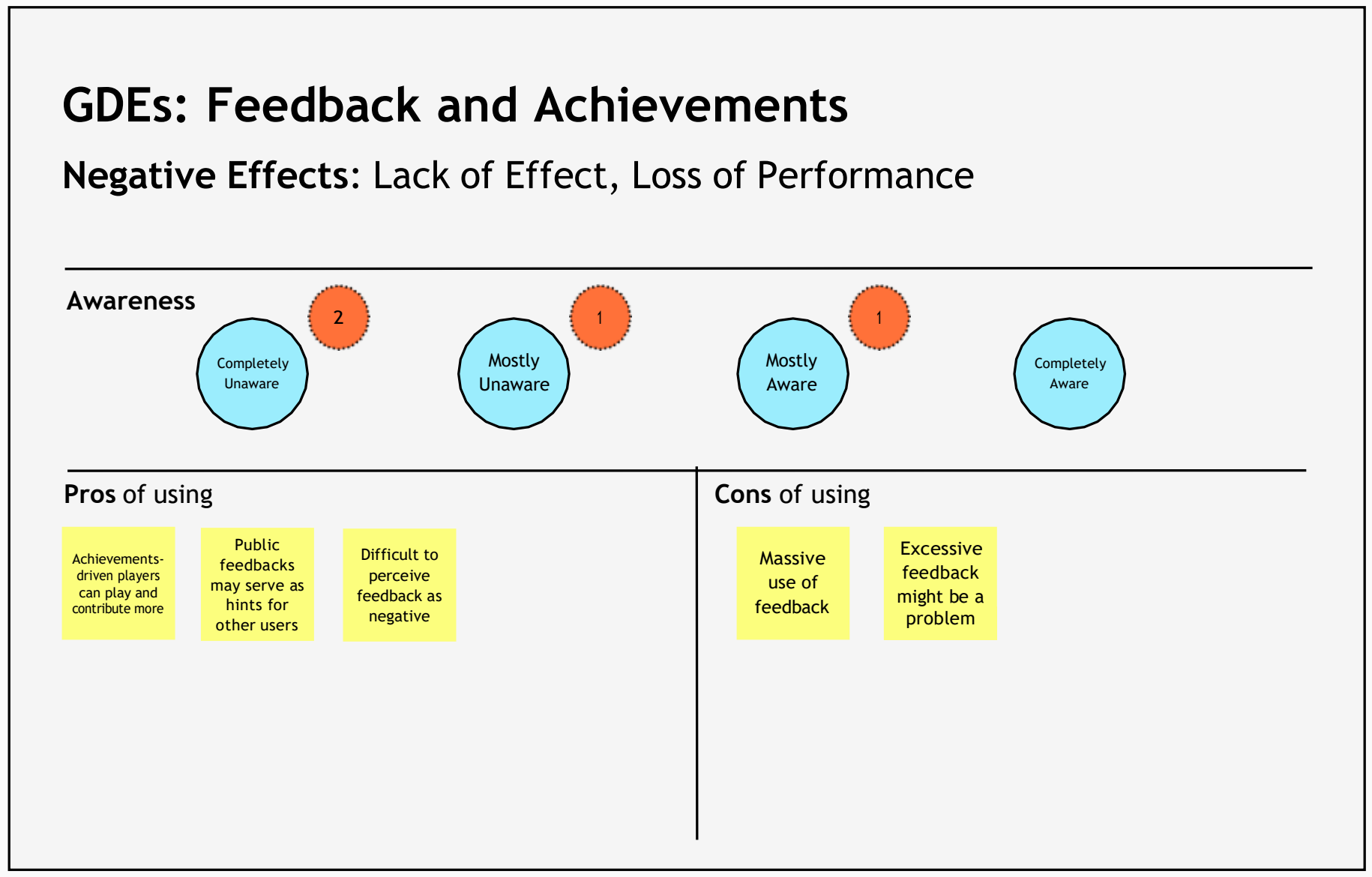

Figure 13: Positive and Negative Comments for Feedback and Achievements.

*P1: "So, there's that question about various gamification elements, but in truth, it has the same meaning. I think, this was our case, in relation to the list of activities that the citizen would be able to do, I don't know if you all agree, but when we say to the other citizens that that player, that citizen, did an activity with success, I believe that it is a valid feedback form to other participants, even as a way to engage. So-and-so is reporting small flies at that region. Hey I'm also going there because theoretically there's a high concentration, or something like that."*

**Difficult to perceive feedback as negative.** Participant P3 mentioned that the use of feedback should be considered in any software. In gamified software , its use is more intensive because when the user performing a task with with reward, the user must always be informed.

*P3: "[...] Feedback is something that must be considered in any software development. [...]. So, as a gamified software this should be even more prominent, because, you start doing a lot of different achievements and you don't get any response from the software that you're changing levels and getting a badge. So it's a little difficult for me to see that lack of effect is a negative feedback effect."*

*Negative comments.* **Excessive feedback might be a problem** and **Massive use of feedback.** The Participant P3 have mentioned that the excessive use of feedback results in a negative and intrusive effect for the user of the software .

> *P3: "Maybe it can be bad when you receive too much feedback. For instance, I logged into the software , and I get a notification, anything you do on the software gets feedback. So maybe there is a negative and intrusive effect."*

### *5.3.6. Avatar*

Figure 14 illustrates the comments made by the participants for Avatars. Three participants were *completely unaware*, and only one participant was *completely aware* about the negative effects. We detail each comment as follows grouped by positive and negative comments.

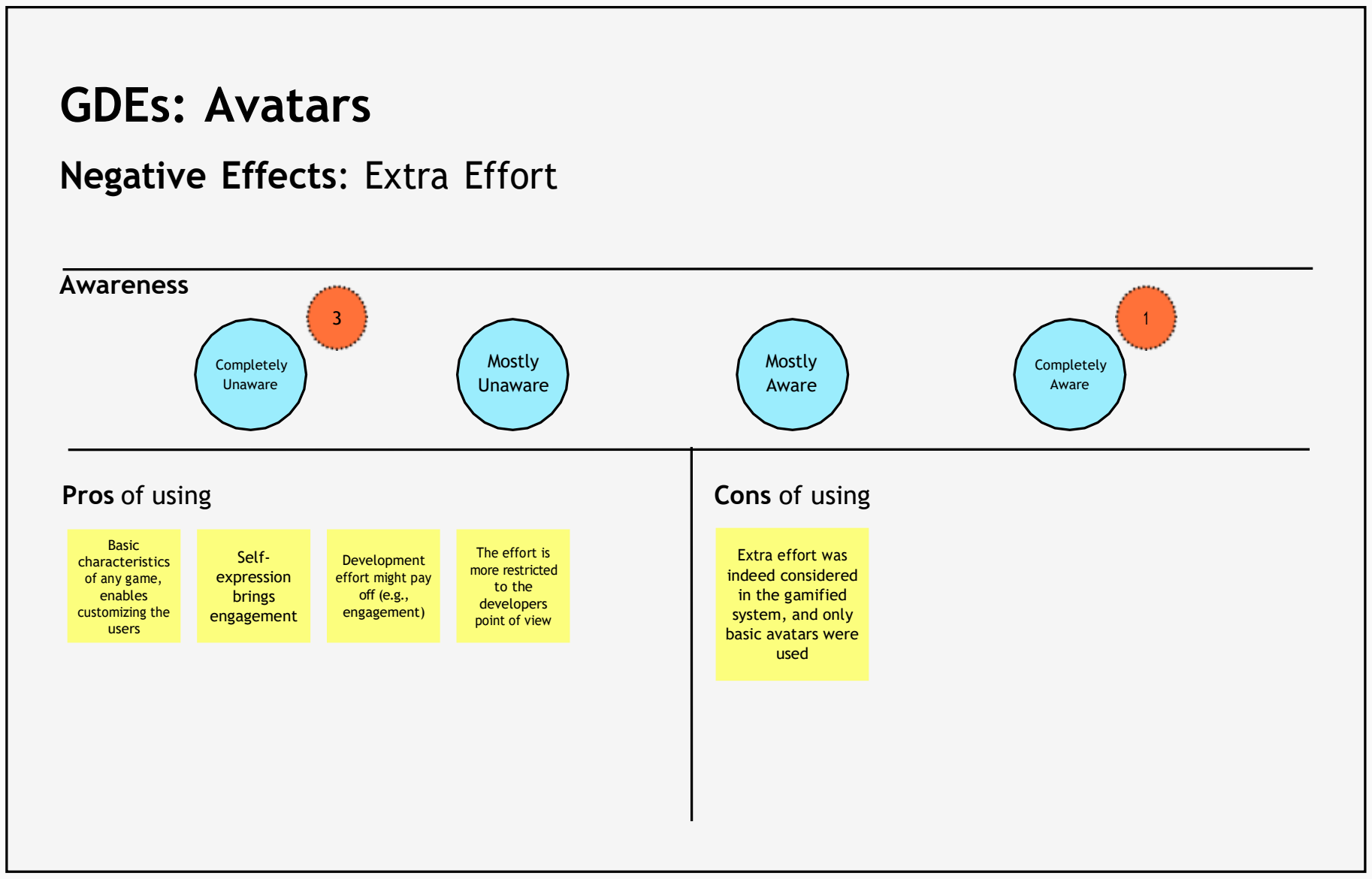

Figure 14: Positive and Negative Comments for Avatars.

*Positive comments.* The participants have mentioned a total of four positive and one negative comments as follows. **Basic characteristics of any game, enables customizing the users.** P1 and P3 have mentioned Avatar is a basic characteristic of any gamified software . Even more so today, in which everyone wants to differentiate themselves, and in this case an avatar can express that difference.

*P3: "In the same way that XP is considered a basic feature of any game, Avatar is too. Even more so nowadays, when everyone wants to differentiate themselves, and you don't have an avatar that can express that difference. For me or at least for most users it is a core feature of the software."*

Additionally, participant P1 mentioned that from the end user's point of view, there is no extra effort in selecting an avatar's characteristics. P1 also mentioned that setting up an avatar is normally part of the initial setup of any game.

*P1: "From the end user's point of view, I don't think it takes too much effort. Also, as P3 mentioned, if you were going to play some other game, or even some gamified software , a step in the tutorial is to create your avatar."*

**Self-expression brings engagement.** The participant P4 have mentioned that in gamified software s that focus on a younger audience, the use of avatars leads to greater engagement, mainly due to the capacity for self-expression.

*P4: "You know, it's a thing that it must exist, because some like it, mainly the young public.[...]"*

**Development effort might pay off (e.g., engagement).** Participant P2 has mentioned that despite the development effort required to add an avatar as a feature into the software , the effort might pay off, especially if the avatar increases user engagement in the software .

*P2: "Understood. Because, even if it demands effort from the developer to create the activity, if me manages to engage (...), even if just 10% or 20% of the users, it would be worthwhile."*

*Negative comment.* The participants P1 and P4 mentioned that an **Extra effort was indeed considered in the gamified software , and only basic avatars were used**. More specifically, P1, mentioned that the creation of sophisticated avatars requires the customization of different items, e.g., hair, clothes, skin color, which consumes development time (one of the main constraints of the VazaZika project).

> *P1: "From the developer's point of view, in VazaZika we didn't have a sophisticated avatar where you could choose your hair, clothes, skin color, etc. One of the comments for this was extra effort, and time restrictions."*

Participant P4 also mentioned that the extra effort is also related to the constant need to make new items available for the avatar, which are often driven by external trends, associated with the need for a professional with design skills.

> *P4: "Who chooses and customizes an Avatar is an identity mechanism, the person will continually update the Avatar. Something happened, I'll update the avatar. Changed a trend, a fashion, I'll update. It's how I change clothes. I update the avatar all the time."*

> *P4: "(...)because it's something that demands much more design, so it must have a designer, someone that would work with that part."*

### *5.3.7. Quizzes*

Figure 15 illustrates the comments made by the participants. Two participants were *mostly unaware*, and two *mostly aware* of the negative effects. We detail each comment as follows grouped by positive and negative comments.

*Positive comment.* The participants mentioned one positive comment: **Might Help To Achieve Educational Purposes** Participant P1 mentioned that, although quizzes were not added to VazaZika software , the quizzes were considered in the early phases of VazaZika software design. The same participant also mentioned that an advantage of using quizzes is enabling achieving educational purposes.

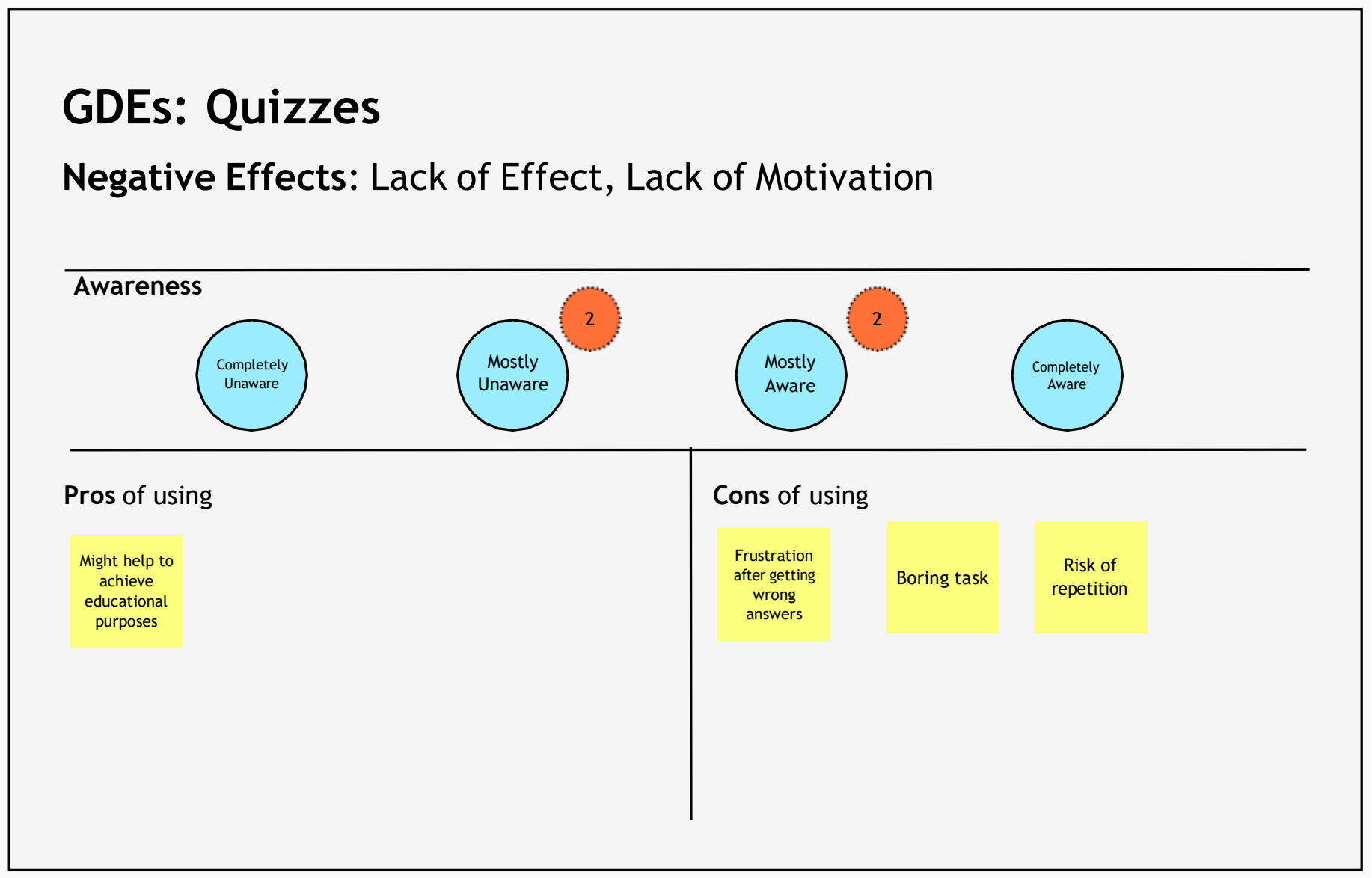

Figure 15: Positive and Negative Comments for Quizzes.

*P1: "I think that the advantage is when you want to gamify in this educational sense. So, in our software the advantage would be that we would be making the citizen be more conscious about the peculiarities of the small fly and its focus. So, how to combat the focus? Which is the time period with more incidence? [...]"*

*Negative comments.* The participants mentioned a total of three negative comments as follows. **Boring task.** P1 mentioned that compared to other gamified tasks, answering questions can demotivate players.

*P1: "[...] only answering questions really is a bit boring, if compared with other tasks that the game would offer."*

**Frustration after getting wrong answers.** In line with P1, participant P3 elaborated on the possible negative effect of using quizzes is the frustration when a user gives a wrong answer to a question, generating pressure to get it right.

*P3: "First that I don't like quizzes, but there are cases where you have to do, right,*

*you don't have a choice. But there's that pressure of having to get it right, and when you don't, you stay with that [...] frustration, and I always see it as a negative effect."*

*P3: "Even more when you are in an educational software, where the person answering the quiz wants to get it right and when it doesn't, or has the pressuure to, or gets frustrated.[...]"*

**Risk of repetition.** The use of traditional quizzes may have the risk of repetition, resulting in the lack of motivation of the players as pointed by P3 as follows.

*P3: "The repetition risk that I added here, its obvious that there are a lot of mechanisms to avoid [it], but a more conventional quiz where you raffle questions in a small group always have that."*

### *5.4. On the Perceived Usefulness, Ease of use and Intent of Adoption of Mapped Negative Effects*

As designed, based on a simplified and direct approach on the main TAM constructs, we asked participants to assign one vote to their *agreement degree* on the usefulness, ease of use, and intent of adoption of information on potential negative effects of game design elements to support adoption decisions. We have collected the agreement degree by the participants in the session's virtual mural (as illustrated in section G of Figure 7) based on a five-point Likert scale [114].

Figure 16 illustrates the level of Agreement reported by the participants.

All participants strongly agree about the *usefulness* of the potential negative effects of game design elements to support adoption decisions. On the other hand, only one participant partially disagrees and three participants partially agree that the information about the potential negative effects are *easy to use*. Finally, one participant partially agrees and three participants strongly agree on the *intent to use*. With regard to these questions, participant P1 mentioned that information about the negative effects are sensitive to the context of the gamified software: *"[...] depending on the context of your software, it may be that all negatives effects become positives, for instance. So, depending on the context,*

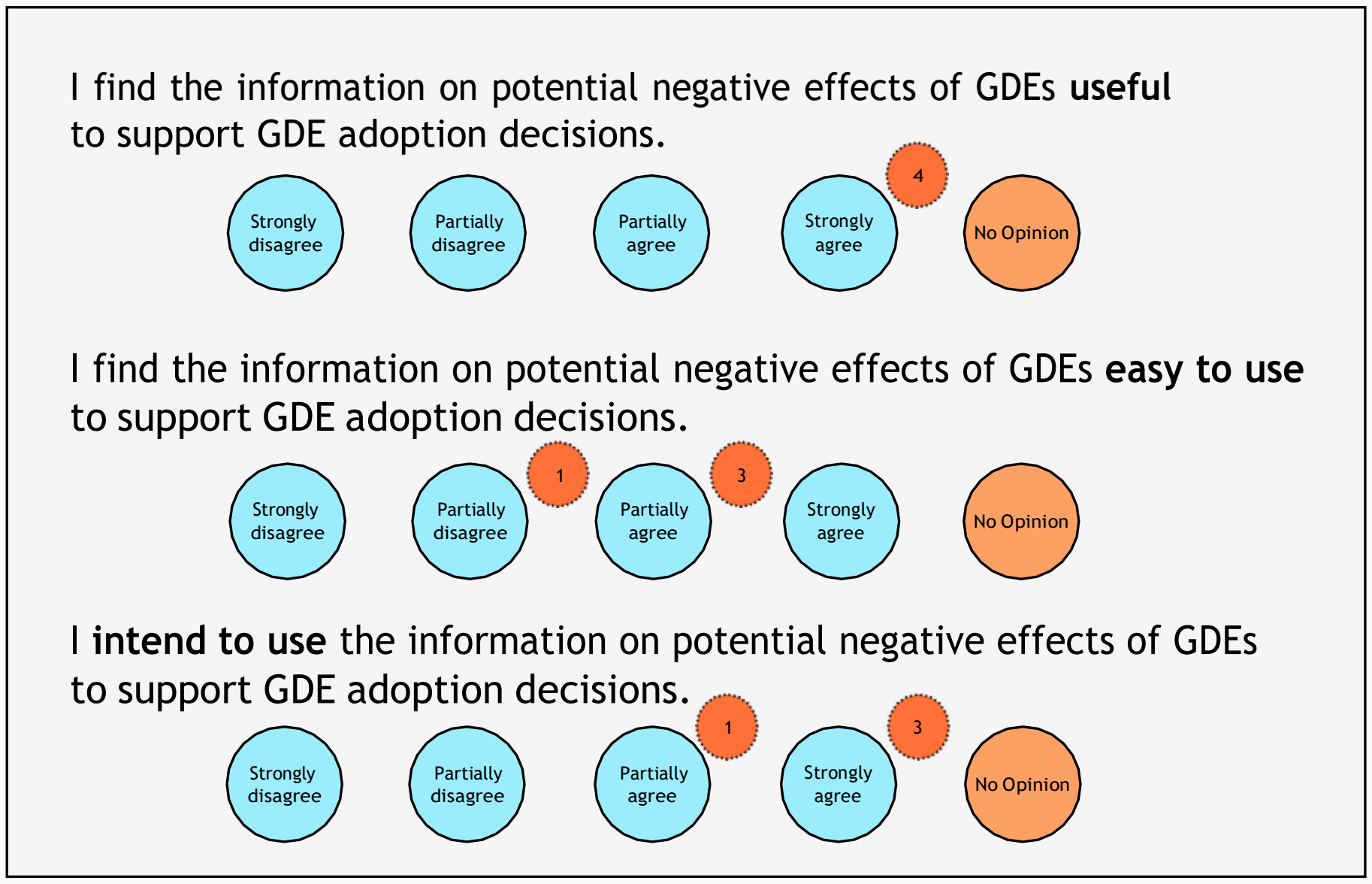

Figure 16: The level of Agreement on Usefulness, Ease to Use, and Intention to Use.

*and if a catalog is created, it is very sensitive to the context of the application."*. Additionally, P2 mentioned that information about negative effects is important for developers who have never implemented a gamified system: *"[...] to learn from other people's experience, for example with a succinct catalog so that you can get information quickly and learn, I think it is extremely valuable."*.

### *5.5. Participant Feedback*

Table 17 summarizes the collected participant feedback. This form was composed of two questions aimed at capturing the degree of confidence and comfort of the participant during the focus group session. All participants agree or strongly agree that they were able to discuss the negative effects of each group of game design elements. Additionally, all participants strongly agree that they felt comfortable in sharing their opinion during the focus group.

Table 17: Participant Feedback Collected via Feedback Form

| Question | P1 | P2 | P3 | P4 |
|---|---|---|---|---|
| 1) I was confident when discussing the negative effects of each group of game design elements | Strongly agree | Agree | Agree | Strongly agree |
| 2) I was comfortable with sharing my opinion during the discussions | Strongly agree | Strongly agree | Strongly agree | Strongly agree |

## 6. Limitations

One of the limitations of our mapping study concerns potentially missing papers. After analyzing 3458 papers (see the list in our online repository), based on our inclusion and exclusion criteria, we initially included 79 papers. While our combined search strategy allowed identifying significantly more papers than the database search strategy employed by Toda et al. [27] (we identified 34 papers published between 2012 and 2016, against 17), the sets had differences. Therefore, we manually included the papers found by Toda et al., which were missed by our first search and extended searches, ending up with a final set of 87 papers.

It is noteworthy that we verified that many of the missed papers would also have been retrieved by subsequent snowballing iterations. Nevertheless, an extension applying subsequent snowballing iterations and investigating different hybrid strategies [15], would require significant additional effort, involving analyzing several thousands of papers which could characterize a different publication instead of an extension. We are confident that our final set of included papers as part of this publication allowed providing an unbiased and meaningful overview of the adverse effects related to GDEs in gamified education software.

Another risk of false negatives concerns the filtering process. We screened all papers considering only titles, abstracts, and keywords, which may not contain sufficient information to decide upon inclusion. We avoided applying EC1 and EC3 during the initial screening to lower this risk, only excluding papers that we had high confidence of not investigating GDE effects (EC1) and reporting negative effects (EC3). In case of any doubt, the paper was left for full text-based assessment. Moreover, the application of the inclusion and exclusion criteria made by the first author was reviewed by the second and fourth authors in meetings. In case of doubt, during the initial screening or full-text-based assessment, discussions were held to reach a consensus.

Furthermore, we chose not to consider grey literature as part of our inclusion criteria. There is the possibility of relevant grey literature that does not have equivalent in non-grey academic papers. On the other hand, even though we did not explicitly evaluate the strength of evidence, the results herein reported are based on peer-reviewed research and backed by empirical studies.

Moreover, while research that reaches negative results is essential because it shows us what does not work [115], there still seems to be a publication bias towards positive outcomes. Research reporting negative results tends to have less scientific interest, fewer citations, and to be less often published [116]. Hence, there may be additional negative results that were not published and which, for that reason, could not be included in our mapping study.

Finally, in our summary of the negative effects of GDEs (*cf.* Figure 4), we included negative effects that may affect different roles (e.g., the students and the maintainers/teachers responsible for configuring the system and keeping it up and running with appropriate educational resources), as we found them reported in the literature. Our goal was to organize evidence regarding the overall adverse effects of GDEs in gamified education software. While we included an overview of the negative effects for different roles (e.g., RQ1.B and RQ1.C), we considered a fine-grained analysis of the effects of each GDEs for each of these roles out of the scope of our intended overview and part of future work.

With regard to the focus group session, we wanted to gather insights on the mapped negative effects of game design elements from the point of view of practitioners. In a virtual focus group, generally, the number of participants is reduced, four being considered appropriate (our case) [117]. However, it is suggested to plan more online focus groups with fewer participants than when conducting face-to-face groups [118]. While we carefully planned our focus group, followed best practices, and carefully conducted our qualitative analyses, a single session is surely a limitation. Therefore, we interpret the focus group results as preliminary complementary discussions with practitioners, avoiding validity claims. Nevertheless, we had rich discussions that nicely complemented our literature study findings. It is noteworthy that these discussions already revealed interesting facts, such as the potential unawareness of developers about the negative effects of GDEs, and some example arguments on the perceived pros and cons of including the GDEs in gamified software.

Still, we are aware that a single focus group is not be enough to reach generalizable findings. Unfortunately, identifying teams with experience in developing gamified software and willing to collaborate with academia is not a trivial task. Therefore, conducting new focus group sessions is part of future work.

## 7. Concluding Remarks

This paper reported a systematic mapping of negative effects of game design elements on education/learning software. Based on data extracted from 87 identified papers, we provided a comprehensive overview with information for software engineers and designers of such software. For instance, we identified the game design elements that have most often been related to adverse effects, the most common negative effects, and the relation between game design elements and negative effects.

The visible trend found from our research is that there has been a research focus on a small subset of the field of negative effects of educational gamified software's GDEs. Badges, Leaderboards, and Competition were the three most mentioned GDEs. Lack of Effect, Worsened Performance, and Demotivation were the three most mentioned negative effects.

Worsened Performance was the negative effect found related to the trio of GDEs mentioned above, which makes some sense. GDEs aren't silver bullets - each one of them has different uses, and different preferences and interpretations by different people. For instance, if someone dislikes competition and competition is the main gamification driver of a gamified software, the software won't be able to motivate that user. A lack of motivation (that can be found related to Badges and Competition) can result in a Worsened Performance related to the user's studies.

Beyond these findings, researchers may use the mapping study results to further analyze specific relations between GDEs and effects by examining the related mapped papers. Researchers could also benefit from the overview of available evidence to identify topics on which more primary studies should be conducted.

Furthermore, we conducted a focus group with experienced developers of gamified software that employ several of the mapped game design elements. The focus group allowed to enrich the discussion of the negative effects identified in the literature from the point of view of practitioners. It also allowed us to observe that practitioners may find the mapping of potential negative effects of game design elements useful in practice. Participants of the focus group were not aware of several of the potential negative effects of game design elements and considered the mapping as a valuable asset to help taking such effects into account in trade-off discussions when selecting game design elements in practice.

### *7.1. Future Research Directions*

Given the trends found by our research, more research focusing on the less explored GDEs and their negative effects is recommended. Beyond that, while this research expands the perception of relations between GDEs and negative effects, it is focused on academic research. Grey literature, although not following the same scientific rigor, could help to uncover terrains that researchers could further investigate.

It would also be interesting to systematically review the mapped papers in-depth, gathering further information about the population of each study, the categories of educational software used (coursewares, classroom aids and learning management systems, etc), context factors related to the empirical studies, and details on their outcomes. This could help to better understand the conditions in which the GDEs may or may not generate those negative effects.

Regarding the focus group, we have focused on the viewpoint of developers in order to provide relevant insights into trade-offs between expected positive effects and potential negative effects of GDEs. The developers' viewpoint helped us to reveal, for instance, that part of the effects reported in the literature is unknown to the developers. Furthermore, the focus group discussions may help developers who want to create gamified software that leverages the positive effects or minimizes the negative ones. However, investigating the viewpoint of students would surely also be interesting. Thus, future work in this direction might complement the revealed insights of our study.

**Acknowledgements**

This study was financed in part by the Coordination for the Improvement of Higher Education Personnel (CAPES - Coordenação de Aperfeiçoamento de Pessoal de N´ıvel Superior), Brazil, and by the National Council for Scientific and Technological Development (CNPq - Conselho Nacional de Desenvolvimento Científico e Tecnológico), Brazil - Finance Code 001.

We would like to thank the focus group participants for the valuable discussion that enriched this paper.